\documentclass[a4paper,twoside,12pt]{article}
 
\usepackage{graphicx}
\usepackage{psfig}
\usepackage{amssymb}

\newcommand{\ZZ}{\mathbb{Z}}

\title{Charge instabilities due to local charge conjugation symmetry
in (2+1)-dimensions}
\author{F.A.Bais\footnote{bais@science.uva.nl} ~and
J.Striet\footnote{jelpers@science.uva.nl}\\[2mm] Institute for
Theoretical Physics \\ University of Amsterdam \\ Valckenierstraat 65
\\1018XE Amsterdam \\ The Netherlands\date{April, 2003}}

\begin{document}
\maketitle

\begin{abstract}
\noindent 
Alice electrodynamics (AED) is a theory of electrodynamics in which
charge conjugation is a local gauge symmetry. In this paper we
investigate a charge instability in alice electrodynamics in
(2+1)-dimensions due to this local charge conjugation. The instability
manifests itself through the creation of a pair of alice fluxes. The
final state is one in which the charge is completely delocalized,
i.e., it is carried as cheshire charge by the flux pair that gets
infinitely separated. We determine the decay rate in terms of the
parameters of the model. The relation of this phenomenon with other
salient features of 2-dimensional compact QED, such as linear
confinement due to instantons/monopoles, is discussed.
\end{abstract}

\section{Introduction}
\begin{figure}[!htb]
\begin{center}
\mbox{\psfig{figure=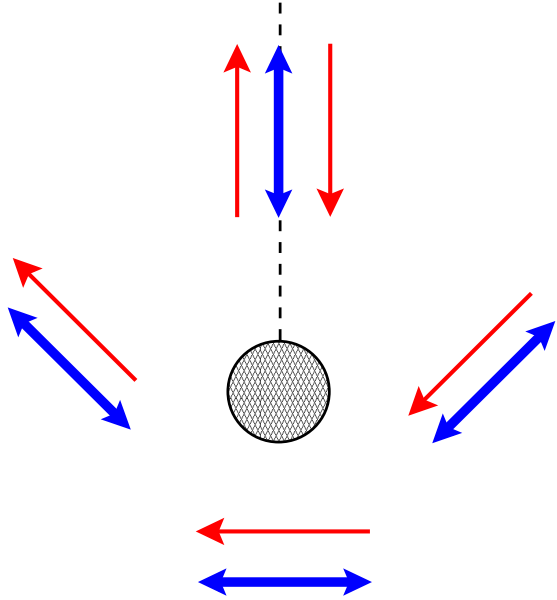,angle=90,width=7.5cm}}
\caption[somethingelse]
{\footnotesize This figure shows the Higgs field, the bidirectional
arrow, configuration of an Alice flux in AED. It also shows that the
generator of the $U(1)$ part of the unbroken gauge group is not single
valued in the presence of such an Alice flux, i.e., a $U(1)$ charge
gets charge conjugated when transported around an Alice loop. }
\label{flux.eps}
\end{center}
\end{figure}

In this paper we investigate charge instabilities in alice
electrodynamics (AED) in (2+1)-dimensions. This theory is closely
related to ordinary electrodynamics. The gauge symmetry of AED is
$U(1)\ltimes\ZZ_2\sim O(2)$, and consists of the $U(1)$ of ordinary
electrodynamics, extended with a local $\ZZ_2$ of charge conjugation,
\cite{schwarz}. In this sense AED is thus a minimally non-abelian
extension of ordinary electrodynamics. However, as this non-abelian
extension is discrete, it only affects electrodynamics through certain
global (topological) features, such as the appearance of alice fluxes,
see figure \ref{flux.eps} (or vortices) and cheshire charges
\cite{alford}\footnote{One might wonder what evidence there is that in
real physics charge conjugation is not a local symmetry, apart from
effects related to those we are about to describe}, see section
\ref{DIPOLE}. Indeed, the topological features of $U(1)\ltimes\ZZ_2$ differs
from that of $U(1)$ in a few subtle but important points. Firstly,
since $\Pi_0(U(1)\ltimes\ZZ_2) = \ZZ_2$, AED allows for topologically
stable vortices, these will be referred to as alice fluxes. Note that
in this theory this localized flux is co\"existing with the unbroken
$U(1)$ of electromagnetism and therefore alice flux is not an
``ordinary'' magnetic flux. If a $U(1)$ charged particle is carried
around an alice flux its charge will be conjugated, see figure
\ref{flux.eps}. This is one of the distinctive features of the alice
fluxes.

Secondly, just as compact $U(1)$ gauge theory, AED contains magnetic
monopoles, because $\Pi_1(U(1)\ltimes\ZZ_2) = \ZZ$.  As is well known
the monopoles become instantons in 2-dimensional electrodynamics and
lead to confinement of charge, see \cite{Belavin} and
\cite{Polyakov2}. The potential between two static charges becomes
linear and the string tension due to the instantons was determined by
Polyakov in \cite{Polyakov2} and is given by:
\begin{equation}
T\propto g^2 \exp{\left(-\frac{S_{inst}}{2 g^2}\right)}\quad,
\label{tension}
\end{equation}
with $S_{inst}$ the action of the instanton in (2+1)-dimensions, or
the mass of a monopole in (3+1)-dimensions, and $g$ the
(dimension-full) coupling constant. In compact alice electrodynamics
there are instantons as well, one therefore in principle expects the
same confining potential between charges. However, as we will see,
whether this confinement will be realized physically depends on the
parameters in the model.

With respect to the monopoles/instantons in AED we have previously
\cite{jelper3} made another observation, namely, that the core
structure of a magnetic monopole may be unstable and deform into a
ring of alice flux carrying a cheshire magnetic charge. This feature,
however is not expected to bear on the confinement mechanism as such,
because the core structure does not affect the long range behavior
of the fields. We return to this point towards the end of the paper.

We see that indeed the topological structure of AED is richer than the
topology of ordinary electrodynamics, as it supports topologically
stable alice fluxes. In this paper we will show that these fluxes may
have a dramatic influence on the infrared behavior of the potential
between two static charges. In the infrared region the potential will
not grow linearly as in ordinary compact electrodynamics, but the
potential will saturate and become constant at a scale set by the mass
of the alice flux. This follows from the fact that a static charge
will be unstable under the creation of two alice fluxes and the
possibility of (induced) cheshire charges carried by such a pair. We
calculate the decay rate of a charge due to this instability, into a
state where the charge is completely delocalized, i.e., virtually
disappeared.

Before turning to a detailed treatment of this remarkable charge
instability, it is useful to briefly discuss some generic features of
the parameter space we are considering. To be as flexible as possible
in separating the various dynamical aspects of the theory, we like to
think of a lattice version of the theory (as discussed in
\cite{jelper4}), because in that setting one can introduce
different mass scales for the fluxes ($m_f$), for the monopoles
($m_m$), and possibly also for dynamical, charged degrees of freedom
($m_q$) by hand. Of course in continuum versions of the model (like
the original $SO(3)$ broken to $U(1)\ltimes\ZZ_2$ model) one often
finds that these physical scales may be linked and one is forced to
restrict oneself to a smaller region of the parameter space then the
one we explore in the remainder of this paper.

The paper is organized as follows. In section \ref{DIPOLE} we examine
the classical configuration of a pair of alice fluxes in the presence
of a charge. We determine the field line pattern of such a
configuration and the energy gain due to the introduction of flux
pair. In section \ref{instability} we analyze the resulting charge
instability in a semi-classical approximation and determine the action
of the bounce solution for some specific decay channels. In the
concluding section we discuss the relevance of our results in the
broader context where one also takes the instantons into account. In
the appendix we introduce the notion of a so called magnetic cheshire
current and point out its relation with electric cheshire charge.

\section{Alice fluxes in the presence of a charge}
\label{DIPOLE}
In this section we examine the classical field configuration due to a
pair of alice fluxes in the presence of a charge. We first analyze
this situation qualitatively, which leads to the conclusion that the
pair of alice fluxes will carry an induced (cheshire) dipole
charge. To see what that looks like we determine the configuration of
electric field lines generated by a conducting needle between two oppositely
charged point charges.  The conducting needle represents a pair of
alice fluxes (one at either end) with their core structure
ignored. Finally, we will determine the energy gain due to the
introduction of the needle/flux pair.

\subsection{The induced cheshire dipole}
\label{dipole}
Let us now study the field configuration of a charge in the presence
of an alice loop (i.e., a flux pair in two dimensions). Due to conservation
and quantization of charge, field lines cannot cross an alice flux, a
situation reminiscent to that of the Meissner effect in a super
conductor. In fact, at first sight one would be tempted to interpret
the whole collection of cheshire phenomena as a manifestation of some
exotic form of electric and/or magnetic super conductivity in the core
of an alice loop. However, this is not possible because the flux tube
itself cannot carry electric/magnetic charge (see also \cite{jelper3})
or current. Let us now consider what happens if we create an alice
loop in the neighborhood of a charge.

\begin{figure}[!htb]
\begin{center}
\makebox[7.9cm]{\psfig{figure=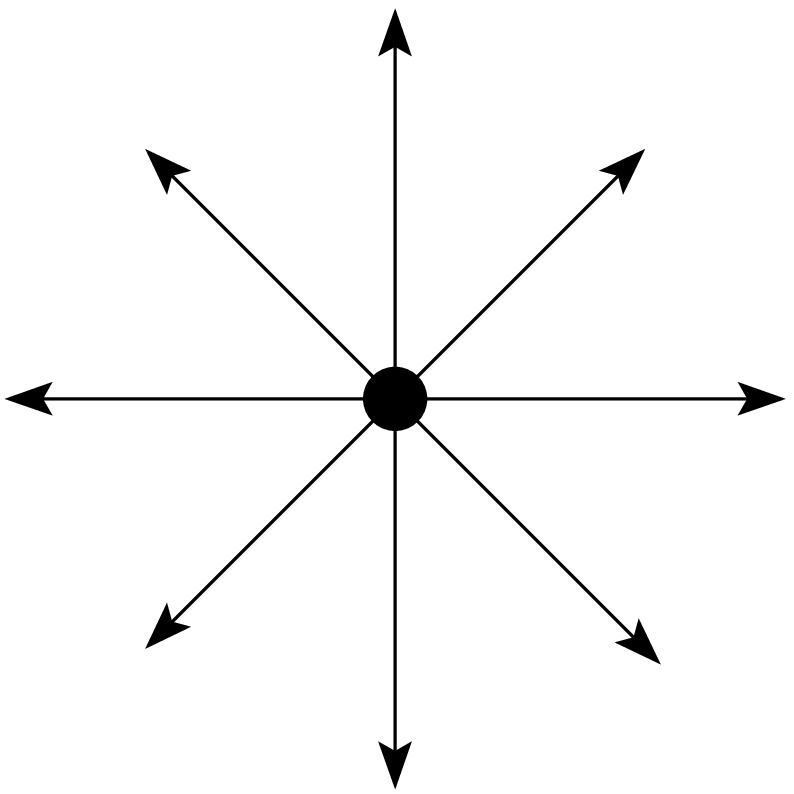,height=3.2cm,angle=0}}
\makebox[7.9cm]{\psfig{figure=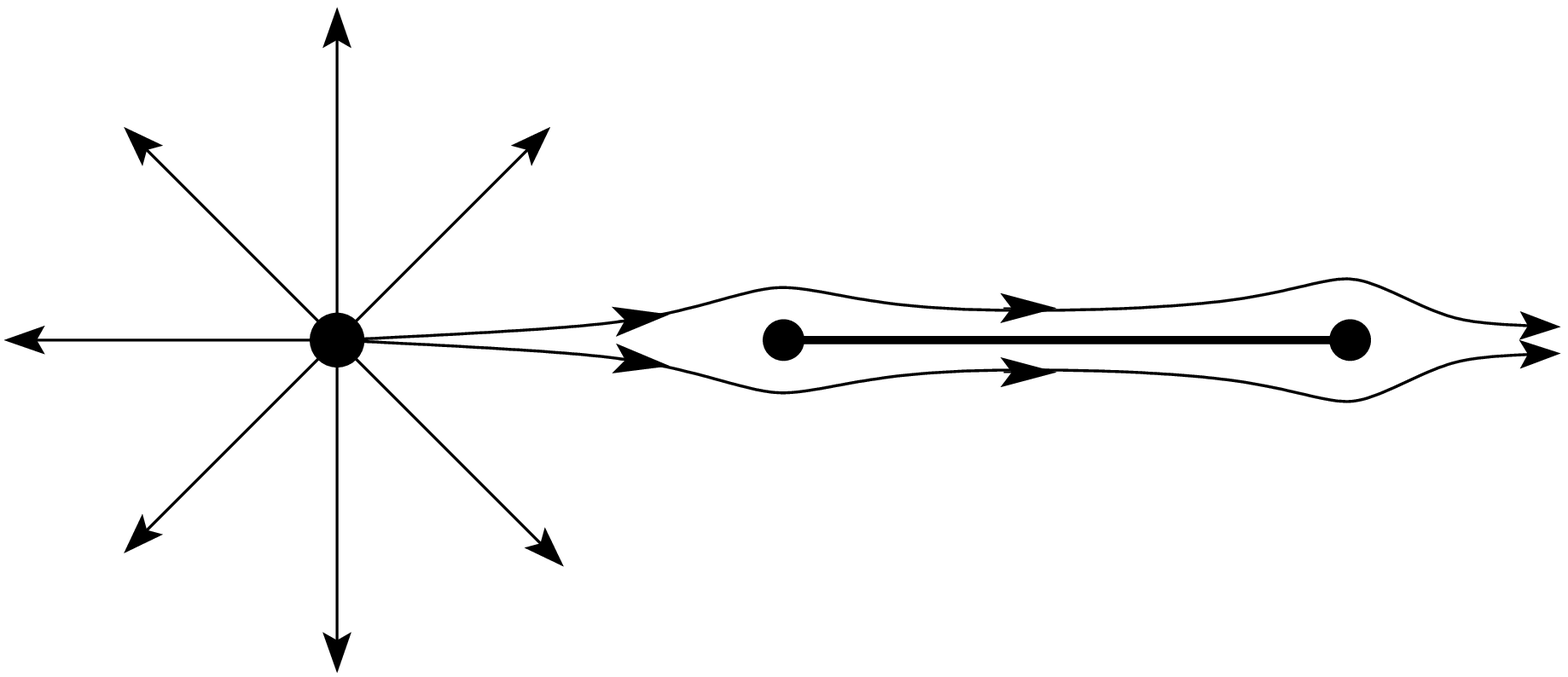,height=3.2cm,angle=0}}
\makebox[7.9cm][c]{\footnotesize{(a)}}
\makebox[7.9cm][c]{\footnotesize{(b)}}
\makebox[7.9cm][l]{\psfig{figure=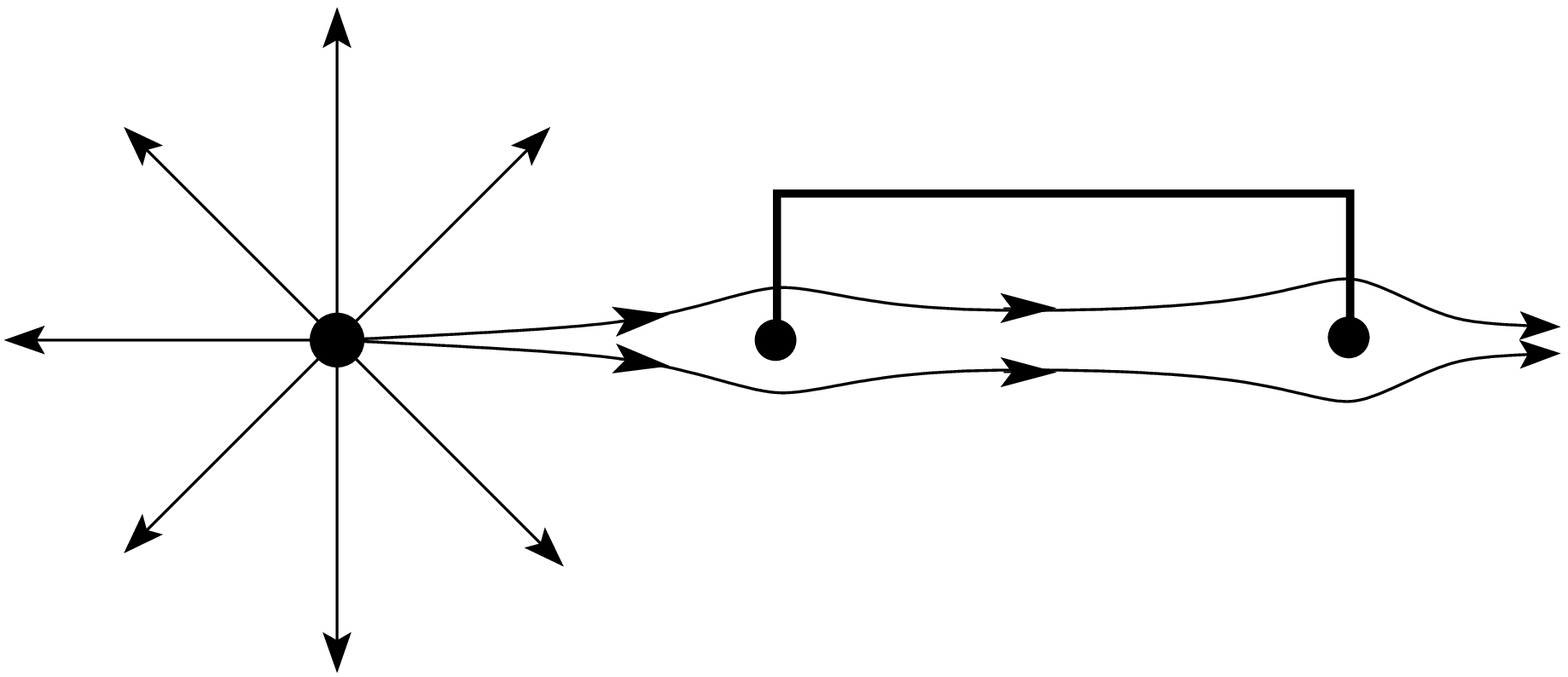,height=3.2cm,angle=0}}
\makebox[7.9cm]{\psfig{figure=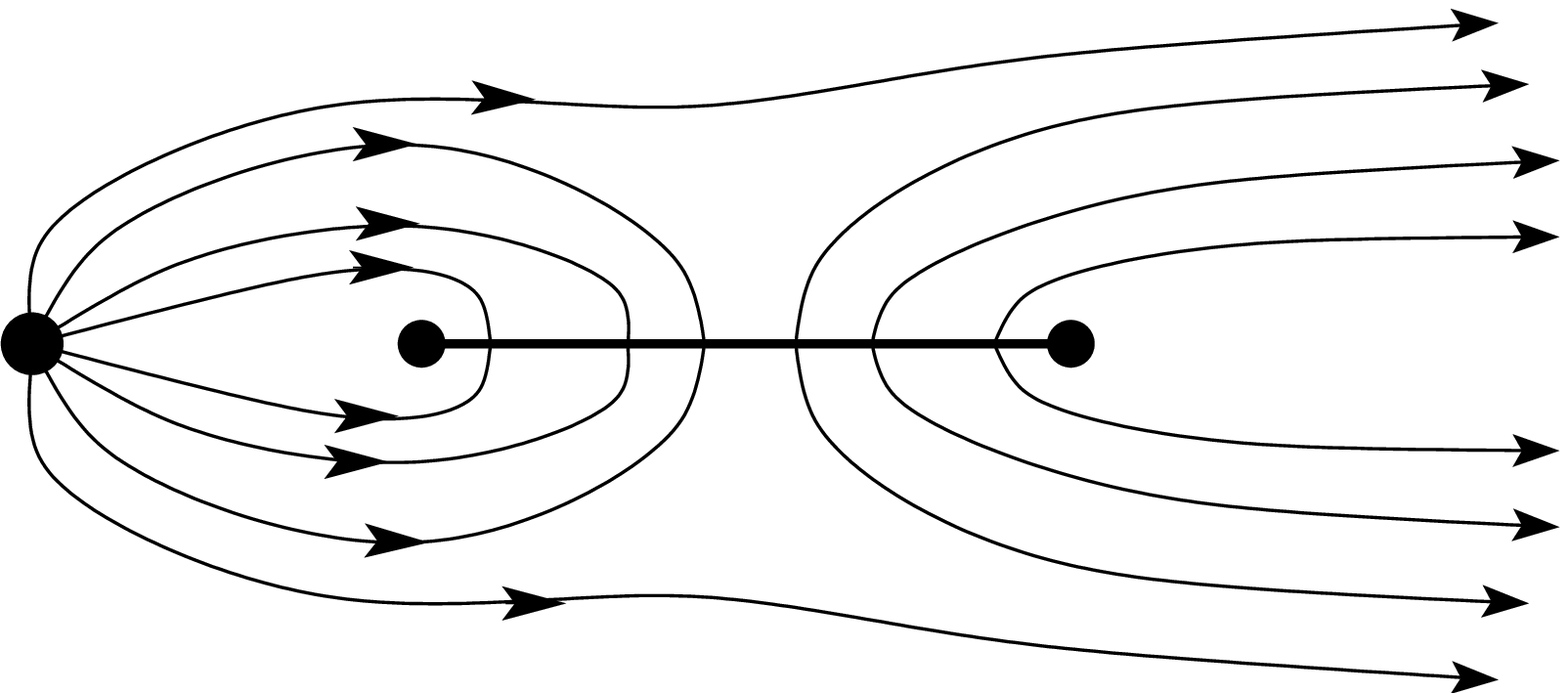,height=3.2cm,angle=0}}
\makebox[7.9cm][c]{\footnotesize{(c)}}
\makebox[7.9cm][c]{\footnotesize{(d)}}
\caption[somethingelse]{\footnotesize\\A sequence of figures that
leads to the correct field line configuration for two alice fluxes in
the presence of a charge. Figure (a) shows a single charge in figure
(b) a pair of fluxes is created in the vicinity of the charge but with
the wrong field line pattern as follows from deforming $\ZZ_2$-gauge
sheet, figure (c). The correct field line pattern is given in figure
(d).}
\label{fieldlines.eps}
\end{center}
\end{figure}

A first guess of how a radial field would be affected due to the
creation of the alice loop might be the same as for the case of a
super conducting loop, i.e., the field lines would be pushed away by
the loop. However the analysis illustrated in figure
\ref{fieldlines.eps} yields a very different picture\footnote{Thus
first one assumes the naively expected configuration to be formed in
analogy with a pair of superconducting wires. However, if one deforms
the $\ZZ_2$-sheet (which is just a gauge artifact) bounded by the
fluxes one sees that that must be wrong, suggesting the correct and
consistent configuration.}. Some of the field lines close around the
first flux while an equal number emanates from the sheet to close
around the second flux and go off to infinity, see figure
\ref{fieldlines.eps}. Thus the total charge carried by the alice flux
configuration stays zero, as it should, but the flux configuration
acquires an induced electric (cheshire) dipole moment. For convenience
we only examine cases where the flux pair lies on the line connecting
the charges. The electric field lines have to be be perpendicular to
the line segment between the two fluxes, because (i) the electric
field lines need to change sign when going around a single flux and
(ii) the reflection symmetry through the horizontal axis of the
configuration.

\noindent In certain symmetric configurations the $\ZZ_2$-sheet may be
considered to act like a conducting plate from which follows that the
charge is pulled towards the alice loop. Indeed, one should be careful
with this analogy because the conducting plate boundary condition of
the $\ZZ_2$-sheet only holds in the particular gauge that satisfies
the obvious symmetry condition. In a general gauge the $\ZZ_2$-sheet
has an arbitrary shape and cannot be interpreted as a conducting
plate. On the other hand, the field line pattern closing partially
around the first and the second flux is gauge invariant (i.e., the
pattern is, but not the direction of the field lines).  We conclude
that the charge induces a dipolar cheshire charge on the alice loop
(or in 2 dimensions, on the pair of fluxes).  This is a natural
generalization of the result obtained in \cite{alford}, but,
straightforward as the generalization may be, there is an important
aspect to it. As we mentioned before, a system of two fluxes or an
alice loop can be in the topologically trivial sector of the theory
and thus may play a role in the dynamical response of the vacuum to an
external charge.

The dipolar behavior of an alice flux pair in the presence of a charge
can have important consequences. Just like a particle anti-particle
pair, these pairs may contribute to the screening of a bare charge,
but an even more drastic consequence is possible. The scenario runs as
follows. One of the fluxes can absorb the point charge, after which
the charge would be carried as a cheshire charge by the flux
pair. This cheshire charge acts like a fictitious charge distribution
along the line connecting the fluxes, generating a repulsive force
between the two fluxes\footnote{We assume for simplicity that a priory
there is no flux-flux interaction. This is not true in general, in the
case of Nielsen-Olesen fluxes it depends on the value Landau
parameter, but if the static forces are zero or repulsive, then the
result obviously holds.}  causing the fluxes to move away from each
other. This would mean that the cheshire charge would increasingly
spread and weaken, put more bluntly, it effectively just disappears.
The fluxes would cause an extreme case of charge delocalization. So,
in two dimensions it therefore appears that in these type of theories,
charge may leak away, implying the absence of any (static) charge.

\subsection{The field configuration}
We now turn to the determination of the field configuration of a flux
pair located between two oppositely charged point particles. We use
the boundary conditions imposed by the fluxes but neglect the core
structure of the fluxes. This boils down to calculating the electric
field configuration of a conducting needle located between two
oppositely charged point particles, where the needle lies on the line
connecting the charges.

\begin{figure}[!htb]
\psfig{figure=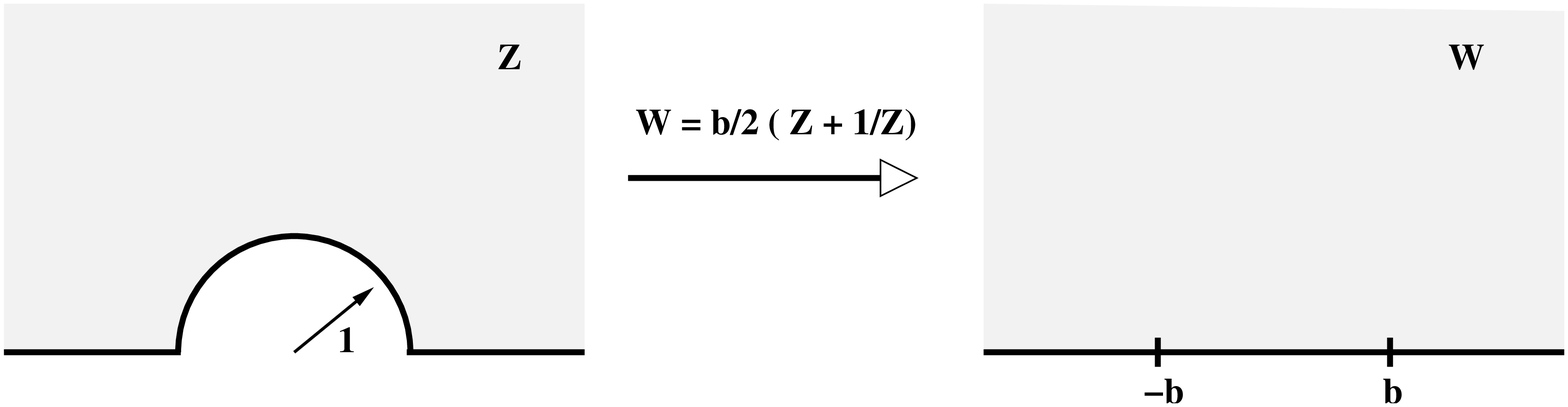,angle=0,width=16cm,height=4.0cm}
\caption[somethingelse]
{ \footnotesize The conformal transformation,
$w=\frac{b}{2}\left(z+\frac{1}{z}\right)$, which maps the conducting disc of
radius one into a conducting needle of length $2b$.}
\label{conftrans.eps}
\end{figure}

Two-dimensional electrostatics (i.e., potential theory) has the
convenient property that it is conformally invariant. Exploiting this
conformal invariance one can construct explicit solutions satisfying
the boundary conditions imposed by the geometry we are interested
in. We start with determining the solution of a charge in the presence
of a conducting disc with the help of the method of images. Then we
use a conformal transformation which maps this conducting disc into a
conducting needle/flux pair, see figure \ref{conftrans.eps}. Since a
conformal transformation is angle preserving, a conductor gets mapped
to a conductor.

\begin{figure}[!htb]
\begin{center}
\mbox{\psfig{figure=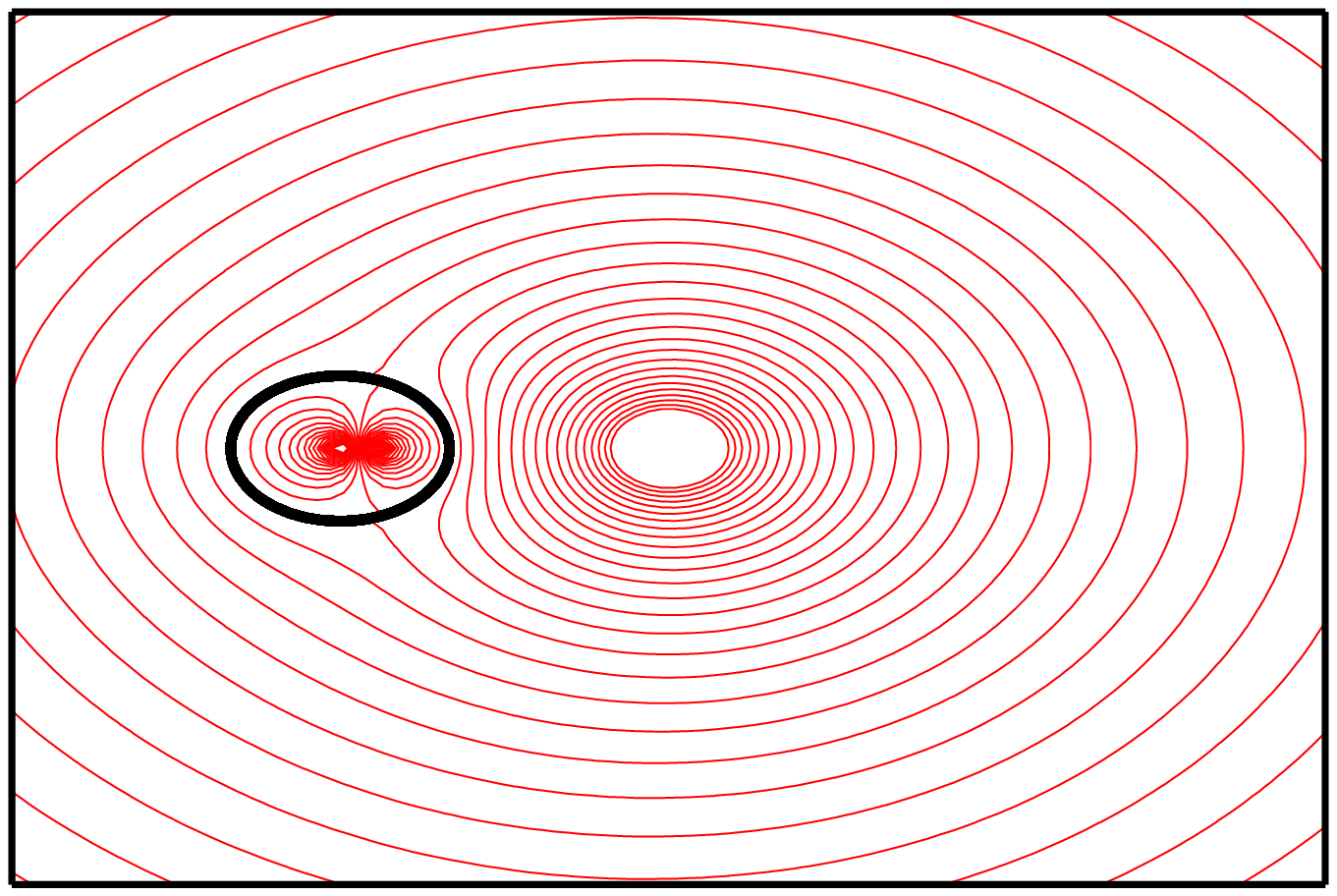,width=7.9cm,height=7.9cm,angle=270}}
\mbox{\psfig{figure=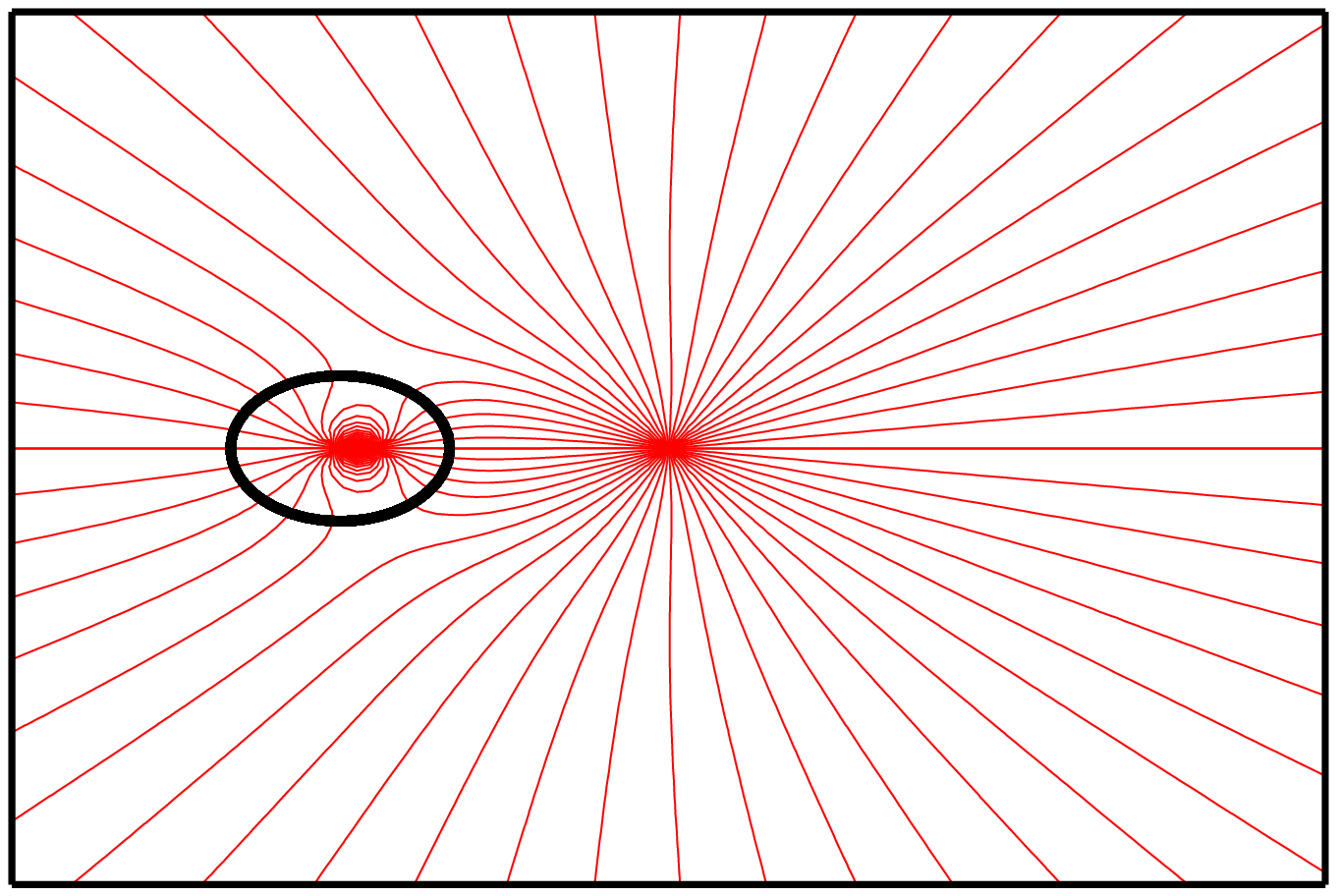,width=7.9cm,height=7.9cm,angle=270}}
\makebox[7.9cm][c]{\footnotesize{(a)}}
\makebox[7.9cm][c]{\footnotesize{(b)}}
\caption[somethingelse]{\footnotesize These figures show some of the
equipotential lines, figure (a), and field lines, figure (b), of a
charge in the presence of a conducting disc. The thick dark circle is
the boundary of the conducting disc. The configuration inside this
circle represents the 'image' charges.}
\label{potdiscc1.eps}
\label{fielddiscc1.eps}
\end{center}
\end{figure}

To construct the configuration of two charges with a
flux pair in between, we first determine the single
charge case and then superpose two of these configurations. We determine
the potential of a charge in the presence of a conducting disc with
the help of the method of images. It is similar to the textbook
example of the charge in the presence of a conducting ball in three
dimensions, but for the case at hand the charge of the
image charges does not depend on the distance of the charge to the
conducting disc. Making use of the identity,
$|\vec{n_1}+a\vec{n_2}|=|a\vec{n_1}+\vec{n_2}|$ with
$|\vec{n_1}|=|\vec{n_2}|=1$, one easily finds the potential $\Phi(z)$,
$z=x+ i y$. The potential is given by:
\begin{equation}
\Phi(z)=\frac{Q}{2\pi}\left\{\log\left|z-z_0\right|-\log\left|z-\frac{R^2}{\left|z_0\right|^2}z_0\right|+\log|z|\right\}\quad,
\end{equation}
with $R$ the radius of the conducting disc, whose center is located in
the origin and $z_0$ denotes the location of the charge. The field
lines correspond with  the height lines of the function:
\begin{equation}
\Psi(z)=\frac{Q}{2\pi}\left\{\arg(z-z_0)-\arg\left(z-\frac{R^2}{|z_0|^2}z_0\right)+\arg(z)\right\}\quad.
\end{equation}
The results are plotted in figures \ref{potdiscc1.eps}a and
\ref{fielddiscc1.eps}b for the equipotential lines and the electric
field lines respectively.

\begin{figure}[!htb]
\begin{center}
\mbox{\psfig{figure=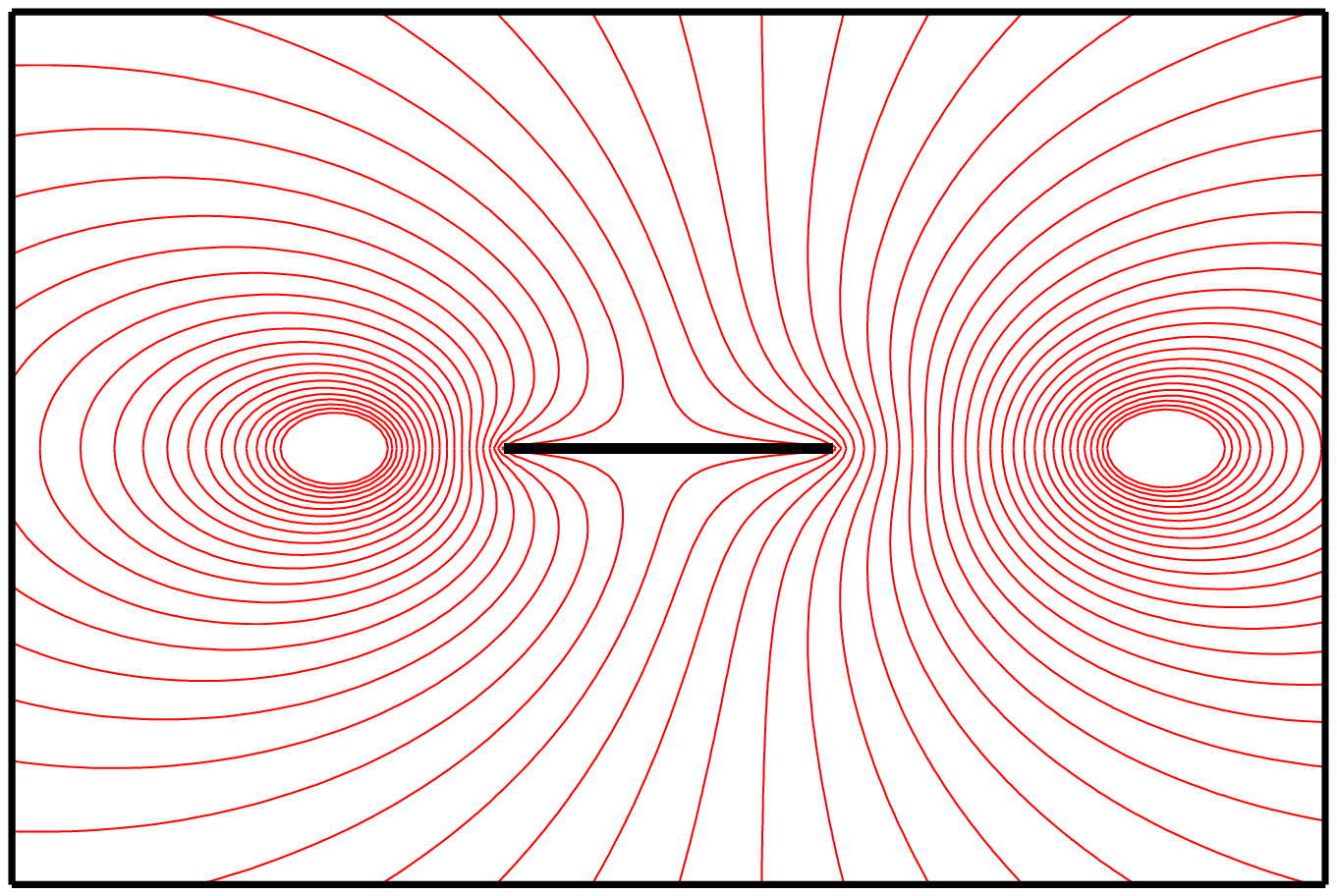,width=7.9cm,height=7.9cm,angle=270}}
\mbox{\psfig{figure=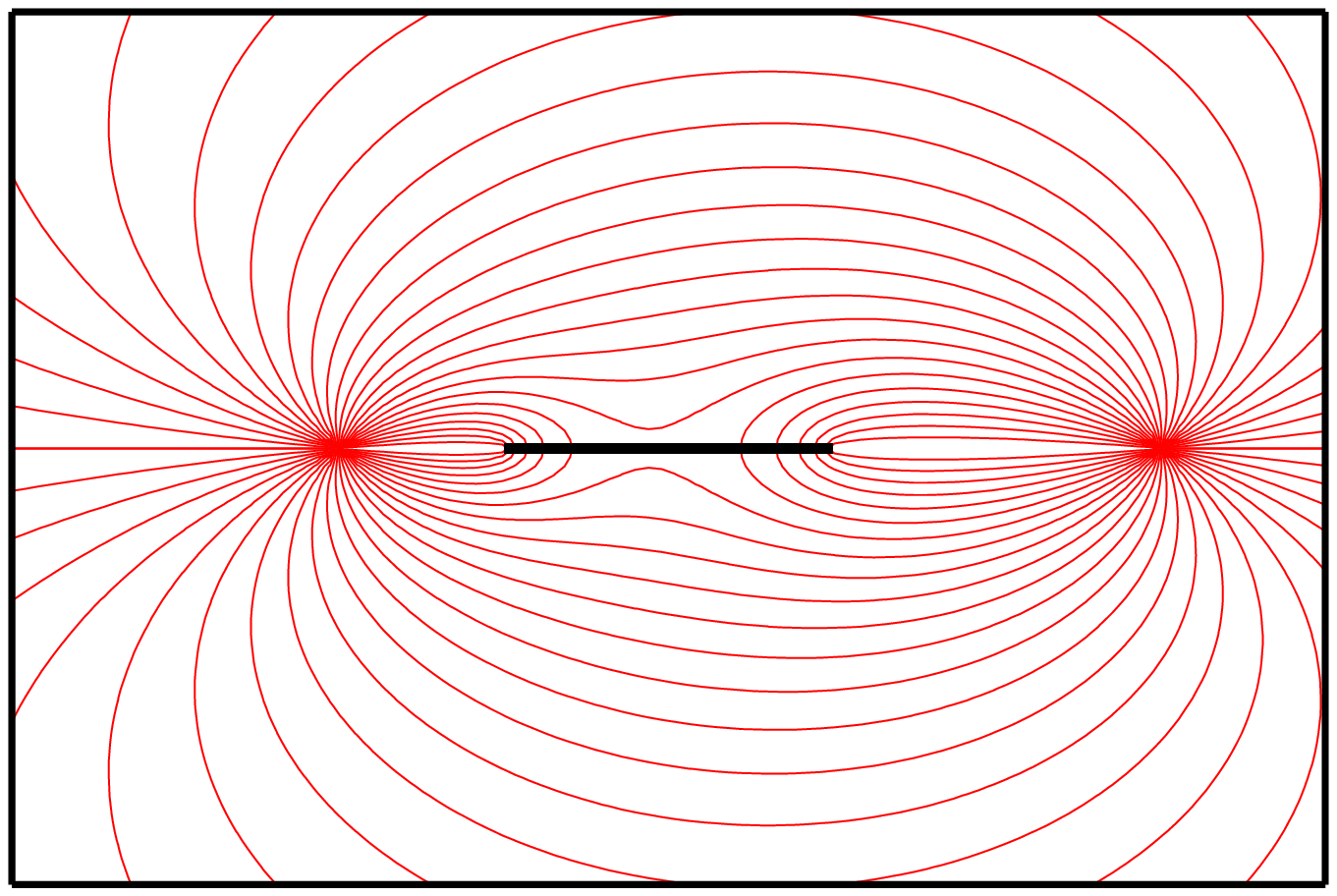,width=7.9cm,height=7.9cm,angle=270}}
\makebox[7.9cm][c]{\footnotesize{(a)}}
\makebox[7.9cm][c]{\footnotesize{(b)}}
\caption[somethingelse]{\footnotesize These figures show some of the
equipotential lines, figure (a), and field lines, figure (b), of two
oppositely charged charges in the presence of the cheshire dipole
carried by a pair of fluxes located at the endpoints of the black line
segment.}
\label{potneedlec2.eps}
\label{fieldneedlec2.eps}
\end{center}
\end{figure}

\noindent We can use this solution to find the solution of a charge in
the presence of a flux pair with the help of the conformal
transformation given in figure \ref{conftrans.eps}. To be more general
we first determine the configuration of two charges in the
presence of a disc. This is straightforward since electrodynamics is
linear in the sense that potentials just add. Thus for the situation
of two (oppositely charged) charges we get the following potential:
\begin{equation}
\Phi(z)=\frac{Q}{2\pi}\left\{\log|z-z_1|-\log\left|z-\frac{R^2}{|z_1|^2}z_1\right| - \log|z-z_2|+\log\left|z-\frac{R^2}{|z_2|^2}z_2\right|\right\}\quad.
\label{phidisc}
\end{equation}
The field lines are now given by the height lines of the function:
\begin{equation}
\Psi(z)=\frac{Q}{2\pi}\left\{\arg(z-z_1)-\arg\left(z-\frac{R^2}{|z_1|^2}z_1\right)-\arg(z-z_2)+\arg\left(z-\frac{R^2}{|z_2|^2}z_2\right)\right\}\quad.
\end{equation}
Let us now use the conformal transformation to map this solution to
the solution of two charges in the presence of a flux pair located on
the line connecting the charges. To be able to use the conformal map,
of figure \ref{conftrans.eps}, $R$ needs to be unity. We can get the
desired configuration if the two charges and the disc also lie on one
line and the disc is between the two charges. We rotate the system
such that $z_1$ and $z_2$ are real. After this we can use the
conformal map to map this solution to the solution of the flux pair
between two oppositely charged point charges. This is done by
replacing $z$ by the corresponding function of $w$, which is given by:
$z=x+\sqrt{x^2-1}$ where we have defined $x = \frac{w}{b}$ and will
use corresponding definitions for $x_1$ and $x_2$. This gives the
following potential:
\begin{eqnarray}
\Phi(x)&=&\frac{Q}{2\pi}\left\{\log\left|x+\sqrt{x^2-1}-x_1-\sqrt{x_1^2-1}\right|\nonumber\right.\\&&-\log\left|x+\sqrt{x^2-1}-\frac{x_1+\sqrt{x_1^2-1}}{\left|x_1+\sqrt{x_1^2-1}\right|^2}\right|\nonumber\\&&-\log\left|x+\sqrt{x^2-1}-x_2-\sqrt{x_2^2-1}\right|\nonumber\\&&\left.+\log\left|x+\sqrt{x^2-1}-\frac{x_2+\sqrt{x_2^2-1}}{\left|x_2+\sqrt{x_2^2-1}\right|^2}\right|\right\}\quad,
\end{eqnarray}
and the field lines follow from:
\begin{eqnarray}
\Psi(x)&=&\frac{Q}{2\pi}\left\{\arg\left(x+\sqrt{x^2-1}-x_1-\sqrt{x_1^2-1}\right)\nonumber\right.\\&&-\arg\left(x+\sqrt{x^2-1}-\frac{x_1+\sqrt{x_1^2-1}}{\left|x_1+\sqrt{x_1^2-1}\right|^2}\right)\nonumber\\&&-\arg\left(x+\sqrt{x^2-1}-x_2-\sqrt{x_2^2-1}\right)\nonumber\\&&\left.+\arg\left(x+\sqrt{x^2-1}-\frac{x_2+\sqrt{x_2^2-1}}{\left|x_2+\sqrt{x_2^2-1}\right|^2}\right)\right\}\quad.
\end{eqnarray}
The conformal transformation only correctly generates the solution in
the upper half plane, $Re(x)>0$. The solution in the lower half plane
follows by the obvious symmetry of the problem.  In figure
\ref{potneedlec2.eps}(a) and \ref{fieldneedlec2.eps}(b) we plotted the
resulting equipotential and  field lines for the configuration.

\subsection{The energy gain}
\label{engain}
In the previous subsection we determined the potential and the field
configuration of a flux pair between two point charges. In this
subsection we calculate the energy difference of this configuration
with the (coulomb type) field configuration without the flux pair. To
be able to determine the energy difference we have to regularize the
expression, i.e., we introduce a UV cut off which will be removed
later. With this cutoff the total energy difference is equal to the
integrated energy density difference. Written in this form, the cutoff
can be removed leaving the energy difference finite, and this is how
we calculate the energy gain due to the presence of the flux pair. To
simplify life the calculation is performed in $z$ space, not in $w$
space. So we use the conformal transformation, which is also just a
convenient change of variables, to transform the solution back into
$z$ space and as an intermediate step, determine the energy gain due
to the presence of a conducting disc and the energy cost due to the
presence of a magnetic super conducting disc. The energy gain due to
the presence of a flux pair is determined from these two results. The
relation between these energy differences is given by:
\begin{eqnarray}
\int\{E_{dipole}-E_{fpair}\}dw =\int\{E_{msc disc}-E_{disc}\}dz\nonumber\\
= \int\{E_{dipole}-E_{disc}\}dz - \int\{E_{dipole}-E_{msc disc}\}dz~,
\end{eqnarray}
where $E_{msc disc}$ is the energy density of two opposite charges
with a disc in the middle, which we identify as a magnetic super
conductor (msc) as the electric field lines are parallel to it. This
configuration is the configuration that one obtains after applying the
inverse conformal transformation, thus from $w$-space to $z$-space, to
the dipole configuration in $w$-space.  First we will determine the
energy gain due to the presence of a conducting disc. This yields the
expression:
\begin{eqnarray}
\Delta E_{disc}&=&2\int_0^\pi\int_R^\infty\left\{(\partial_r \Phi_2(r,\theta))^2+\left(\frac{1}{r}\partial_\theta \Phi_2(r,\theta)\right)^2\nonumber\right.\\&&\left.-~(\partial_r \Phi_1(r,\theta))^2-\left(\frac{1}{r}\partial_\theta \Phi_1(r,\theta)\right)^2   \right\}r~dr d\theta\nonumber\\&& + ~2\int_0^\pi\int_0^R\left\{(\partial_r \Phi_2(r,\theta))^2+\left(\frac{1}{r}\partial_\theta \Phi_2(r,\theta)\right)^2\right\}r~dr d\theta\quad,
\end{eqnarray}
with $\Phi_1(r,\theta)$ given by formula \ref{phidisc}
and $\Phi_2(r,\theta)$ is given by formula \ref{phidisc} with $R=0$. This
gives:
\begin{equation}
\Delta E_{disc} = -\frac{Q^2}{2\pi} \log\left(\frac{(z_1^2-R^2)(z_2^2-R^2)}{(z_1z_2+R^2)^2}\right)\quad.
\end{equation}
The energy gain due to the presence of a magnetically super conducting
(msc) disc is determined by:
\begin{eqnarray}
\Delta E_{mscdisc}&=&2\int_0^\pi\int_R^\infty\left\{(\partial_r \Phi_2(r,\theta))^2+\left(\frac{1}{r}\partial_\theta \Phi_2(r,\theta)\right)^2\nonumber\right.\\&&\left.-~(\partial_r \Phi_3(r,\theta))^2-\left(\frac{1}{r}\partial_\theta \Phi_3(r,\theta)\right)^2   \right\}r~dr d\theta\nonumber\\&& + ~2\int_0^\pi\int_0^R\left\{(\partial_r \Phi_2(r,\theta))^2+\left(\frac{1}{r}\partial_\theta \Phi_2(r,\theta)\right)^2\right\}r~dr d\theta\quad,
\end{eqnarray}
with $\Phi_1(r,\theta)$ given by formula \ref{phidisc} and
$\Phi_3(r,\theta)$ by:
\begin{equation}
\Phi_3(z)=\frac{Q}{2\pi}\left\{\log\left|\left(z+\frac{1}{z}\right)-\left(z_1+\frac{1}{z_1}\right)\right| -\log\left|\left(z+\frac{1}{z}\right)-\left(z_2+\frac{1}{z_2}\right)\right|\right\}\quad.
\end{equation}
One obtains:
\begin{equation}
\Delta E_{mscdisc} = \frac{Q^2}{2\pi} \log\left(\frac{(z_1^2-R^2)(z_2^2-R^2)}{(z_1z_2+R^2)^2}\right)\quad.
\end{equation}
For the case of $R=1$ we have $E_{fpair}=\Delta E_{disc}-\Delta E_{mscdisc}$. Thus we get:
\begin{equation}
E_{fpair} = -\frac{Q^2}{\pi} \log\left(\frac{(z_1^2-1)(z_2^2-1)}{(z_1z_2+1)^2}\right)\quad.
\end{equation}
This result is still in $z$ language, i.e., $z_1$ and $z_2$ need to be
written in terms of $w_1$ and $w_2$. This is done with the help of the
conformal transformation, $z=x+\sqrt{x^2-1}$, and leads to the
following expression for the energy gain:
\begin{equation}
E_{fpair} = \frac{Q^2}{\pi} \log\left(\frac{1}{2}\left(1+\frac{1+x_1x_2}{\sqrt{x_1^2-1}\sqrt{x_2^2-1}}\right)\right)\quad.
\label{Eneedle}
\end{equation}
We see that the energy gain due to creating a flux pair
between two charges is basically unbounded. Moving the flux pair closer
to one or both of the charges increases the energy gain. One expects
that due the renormalization of the charge this would not go on for
ever, effectively one expects an effective UV cutoff.

Let us now investigate the single charge configuration, i.e., we send
one of the charges to infinity. In this case the energy gain is given
by:
\begin{equation}
E_{fpair} = - \frac{Q^2}{\pi} \log\left(\frac{4\sqrt{d}}{\left(1+\sqrt{d}\right)^2}\right)\quad,
\label{EneedleAsymm}
\end{equation}
where $d$ is the ratio of the distance of the two fluxes to the
charge.\\ We find that the energy gain due to the presence of the flux
pair only depends on the ratio of the distance of the two edges to the
charge. Thus no matter what the size is of the UV cutoff, the flux
radius or in fact any other length scale, the energy gain can always
be as large as one wants in a region where all length scales are
insignificant with respect to the distances of the fluxes to the
charge and between the fluxes. This shows that in two dimensions a
single charge is always unstable (or meta-stable) with respect to a
decay into a flux pair with a cheshire charge no matter what the
length scales are. However, the length scales of course drastically
change the decay time of a charge.

\section{The charge instability}
\label{instability}
In this section we analyze a novel type of instability in the electric
field of a charge. We pointed out before, that a pair of alice fluxes
in the presence of a charge acquires an induced dipole, subsequently
we determined the energy gain due to the creation of such pair. This
raises the question to what extend the electric field configuration of
a pair of static localized charges remains stable with respect to flux
pair creation.  We study this question in a lattice version of AED
(LAED). The reason is, as mentioned in the introduction, that LAED
allows us to introduce independent parameters, a mass $m_f$ for the
alice flux and a mass/action $m_m$ for the monopole/instanton. First
we analyze the charge instability, then we will determine what the
decay time is and compare it with the instability under the creation
of a pair of charged point particles (with mass $m_q$), assuming that
these are present in the theory. To what extend these results can be
carried over to a continuum version of the theory will be discussed in
the concluding section.

Before turning to the the detailed calculations, let us make some general
observations concerning the role of the various mass scales in the model.
If both $m_m$ and $m_f$ are very large, a charge in two dimensions 
generates the well known logarithmic potential in the classical (small
$g^2$) limit.: 
\begin{equation}
V(r)=\frac{Q^2}{\pi}\log\left(\frac{r}{r_0}\right)\quad,
\end{equation}
with $r_0$ some UV cutoff. Needless to say that the presence of
dynamical charges in the model would (a) give rise to the standard
(short distance) renormalization of the charge and (b) provide a
cutoff to the potential at an energy of the order of mass of the
charged particles $m_q$. If the monopole mass $m_m$ comes down and
$m_f$ remains very large we get that the monopoles cause confinement,
i.e., a linearly rising potential and the role of dynamical charges
would be very much the same as for the logarithmic case. For the
moment however, we will assume that no charged dynamical particles are
present in the model (i.e., we assume them to be very massive). If now
the flux mass comes down as well, then of course we get the
possibility to dynamically create flux pairs out of the vacuum and
these will cause the decay of the electric fields generated by the
external charges. One expects a situation to arise where the potential
(irrespective of its character) basically saturates and turns into a
constant at a distance $(r/r_0)$ where fieldenergy becomes comparable
to the value $2m_f$.

\subsection{The life time of charge}
Let us now compute the decay time of a system of two charges by
performing an instanton calculation in the spirit of the ``false
vacuum'' as described by Coleman and Callan \cite{Coleman},
\cite{Callan}. To lowest order in $\hbar$ one only needs to determine
the bounce solution with lowest action. The bounce is a classical
solution of the Euclidean system, i.e., with the original potential
inverted. In the mechanical analogue a classical particle moves from
the meta stable point to the corresponding point at the other side of
the barrier and back again. The instability, i.e., the tunneling
through the barrier corresponds to half the Euclidean bounce solution,
after which a real Minkovski time evolution takes over. At this point
the system is not yet in its final state, but one expects that the new
lowest energy state will be reached by emitting/dissipating energy
through conventional (in this model presumably primarily
electromagnetic) radiation processes. In the mechanical system with
the inverted potential one should then find the particle trajectory
with minimal action $S_b$. In the semi-classical domain the decay time
is given by:
\begin{equation}
\tau\propto e^{\frac{S_b}{\hbar}}
\end{equation}
 In our system we find
two extremal paths. We expect one of these two to have
the lowest action, independent of the distance $2w$ between the two external
charges. In the following we analyze the situation for two cases,
firstly we will determine the action for the instability due to the
creation of a flux pair, then we do the same for the creation of
a pair of point charges and finally we compare both mechanisms.

We first consider the case where the pair of fluxes or of charges are
created in the most symmetric way. This means that they start out
exactly between the external charges. The other decay channel we
investigate corresponds to the most asymmetric configuration, where
the fluxes or charges are created in the vicinity of one of the
external charges and only one flux or charge will move. The other flux
or charge remains with the charge at a fixed minimal distance $R_0$,
which represents the UV cutoff of the bare charge. We will also
determine the action of the bounce - the pair creation rate - in a
constant electric field.

So the calculations we are about to make for the various cases are
very similar, so let us, before providing the specific details for
each case, give the general structure of the results.

In the previous sections we have calculated the energy gain $E$ in the
electric field due to the pair creation. From that we can determine the
potential $V_{pair}$ for the creation of a pair as a function of their
separation $2b$ and of course also dependent on the other fixed
parameters that characterize the configuration, such as the external
charges $Q$, their separation $w$, the masses $m_f$ $($or $m_q)$ and
sometimes a core size $R_0$.

\begin{figure}[!htb]
\begin{center}
\mbox{\psfig{figure=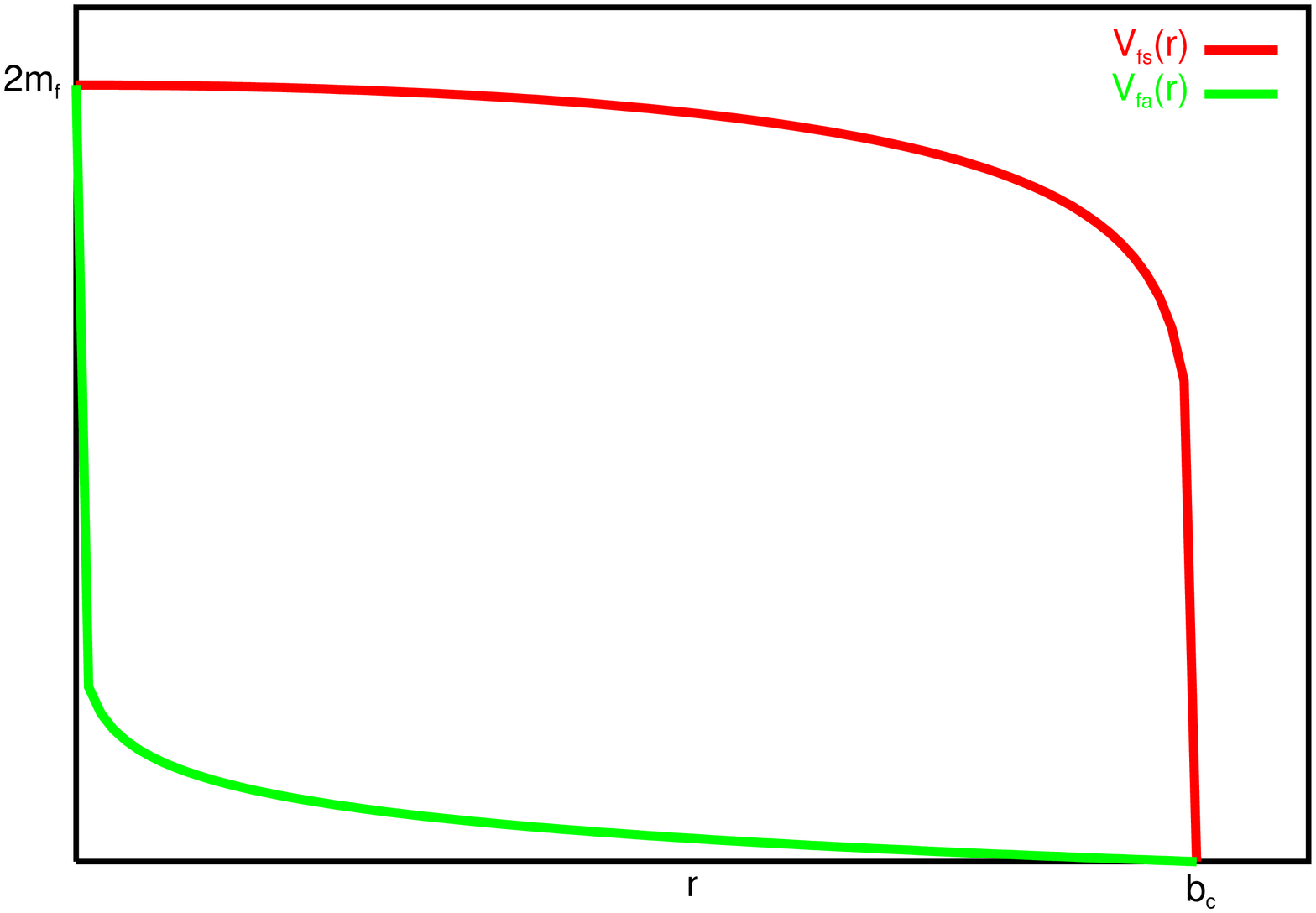,width=7.9cm,angle=270}}
\mbox{\psfig{figure=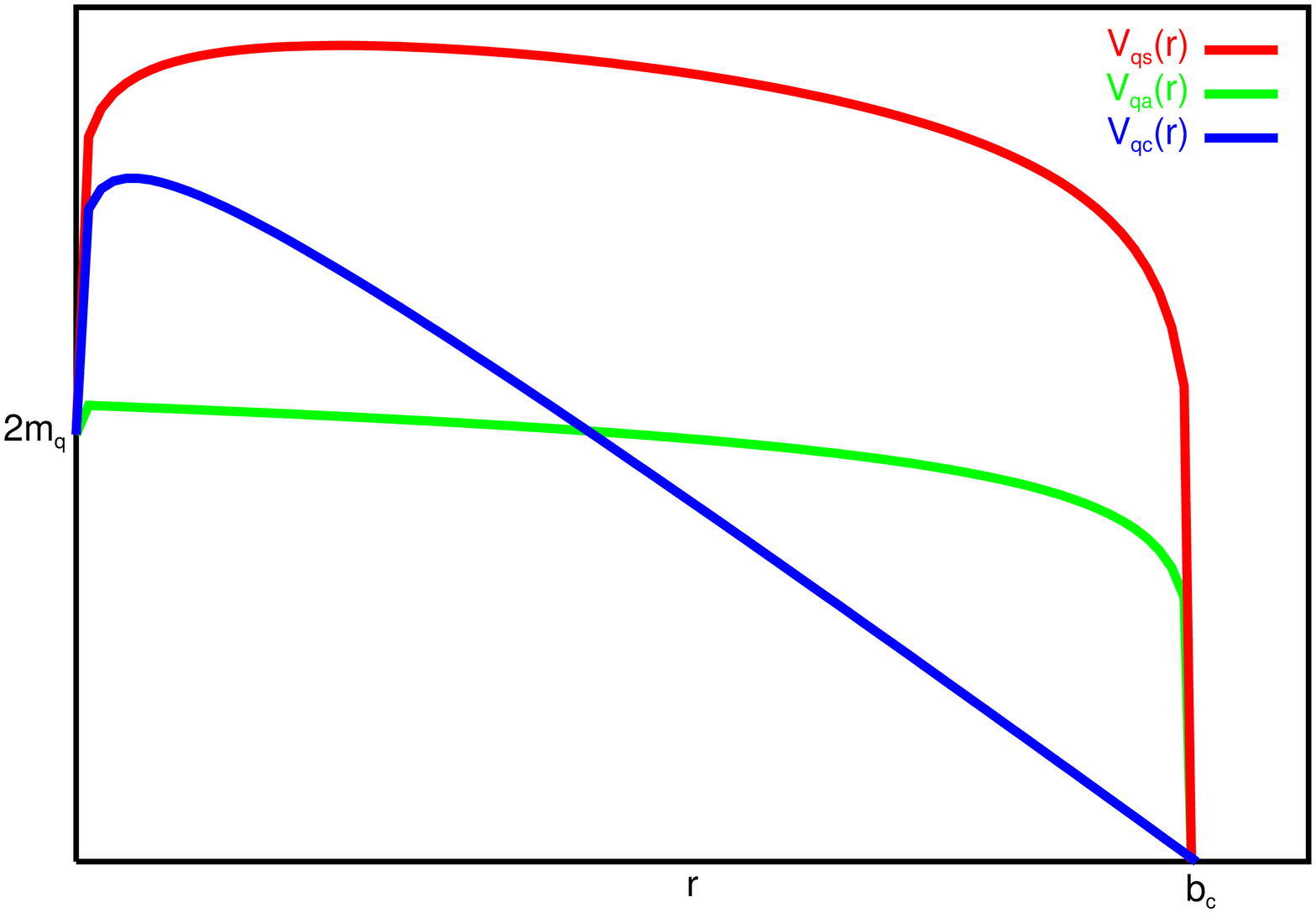,width=7.9cm,angle=270}}
\makebox[7.9cm][c]{\footnotesize{(a)}}
\makebox[7.9cm][c]{\footnotesize{(b)}}
\caption[somethingelse]{\footnotesize In figure (a) we plotted two
typical potentials for the bounce of two alice fluxes in the
symmetric and asymmetric channel respectively. In figure (b) we
plotted two typical potentials for the bounce of two dynamical
charges in the symmetric and asymmetric channel respectively and the
potential for the bounce of two dynamical charges in a constant
field.}
\label{fluxpot.eps}
\label{charpot.eps}
\end{center}
\end{figure}

We have indicated the generic shape of the potentials in figures
\ref{fluxpot.eps}(a) and \ref{charpot.eps}(b) for the pair creation of
fluxes and dynamical charges respectively. For the fluxes we have
assumed there to be no flux-flux interactions so that only the mass
$2m_f$ comes in. For the charged pair, however one expects the
potential to grow with separation which means that the maximum of the
potential is shifted towards larger separation. As is well known in
one dimensional physics, the action of the extremal path generically
is given by:
\begin{equation}
S^{pair} = 2 \int_{b=0}^{b=b_c} \sqrt{4m V_{pair}}~db
\label{SSymNeedle}
\end{equation}
We can bring this expression in a more or less canonical form. One
first introduces a dimensionless separation variable $y$ obtained by
conveniently scaling $b$ with some relevant length scale, for example
the critical separation $b_c$ labeling the turning point, this brings
out a factor of the relevant length scale out in front. Next one
scales the potential by its maximal value: $V= V_{max} \hat{V}$.
$V_{max}$ may conveniently be written as $V_{max} = 2m \gamma^2$ where
$\gamma$ is a dimensionless quantity satisfying $\gamma \geq 1$ and
the equal sign applies to the flux pair creation (see
figures). Putting the scaling factors in front of the integral the
expression for the action takes the general form,
\begin{equation}
S^{pair} = const.\times b_c~ m~\gamma~ F(w,m,Q,R_0)
\label{genaction}
\end{equation}
where the dimensionless function $F$ may depend on all the parameters,
but, because of the rescalings, takes on only values between zero and
one.
\begin{equation}
F= \int_{y=0}^{y=1} \sqrt{\hat{V}_{pair}}~dy
\label{genfunction}
\end{equation}
We see that the action is typically of the order (mass of
pair)x(critical separation), as one would expect naively. Yet, we will
study the various cases separately in more detail, because it turns
out that there are interesting differences in the functional
dependence of $S^{pair}$ on for example the distance $w$ of the
external charges, which are important physically.

\subsubsection{Charge decay due to creation of an alice flux pair}
We compute the action for a bounce corresponding with the creation of
a flux pair in the presence of two external charges. First we consider
the symmetric channel, then the asymmetric channel and finally the
case of a constant electric field.

 \noindent\textbf{The symmetric channel:}

In the symmetric channel we may use formula
\ref{Eneedle} with $x_1=x_2=x$, which gives the energy gain:
\begin{equation}
E_{fpair} = -\frac{Q^2}{\pi} \log\left(1-\frac{1}{x^2}\right)\quad.
\label{SymEfpair} \label{SymENeedle}
\end{equation}
During the bounce the external charges remain fixed while the distance
between the fluxes increases. The suitably scaled variable for this
situation is $y \equiv \frac{1}{x} = \frac{b}{w}$.  So far we only
determined the energy gain due to the boundary conditions created by
the alice fluxes, but the potential in which the fluxes move is not
only given by the energy gain, we also should include the energy cost
which equals the mass of the flux pair, $2m_f$. The potential for the
pair is therefore given by:
\begin{equation}
V_{fpair} = 2m_f \left(\Theta\left(|y|\right)+ \frac{1}{\mu} \log\left(1-y^2\right)\right)\quad,
\label{SymVNeedle}
\end{equation}
where $\Theta(0)=0$ and equals one otherwise. The constant $\mu$ is
defined as $\mu=\frac{2\pi m_f}{Q^2}$. We should note that keeping
$y_1=y_2=y$ for all times is in fact a solution to the equations of
motion for the system with the inverted potential. The action of this
solution is simply given by:
\begin{equation}
S^f_{sym} =  4\sqrt{2}~ b_c~ m_f F^f_{sym}(\mu)
\end{equation}
where the turning points are given by the zeros of the potential, i.e.
\begin{equation}
b_c = w~\sqrt{1-e^{-\mu}}
\end{equation}
and where $F^f_{sym}(\mu)$ is given by:
\begin{equation}
F^f_{sym}(\mu) = \int_0^1 \sqrt{1+\frac{1}{\mu}\log\left(1-y'^2\left(1-e^{-\mu}\right)\right)}~dy'\quad.
\end{equation}
Note that the function $F^f_{sym}$ depends in this case only on
one particular combination of parameters, $\mu$. The integrand varies
from one at $y=0$, to zero at $y=1$. Although the integral cannot be
done analytically, a little analysis shows that the function always
lies between the functions $f(y)=\sqrt{1-y^2}$ and $f(y)=1$. The
integrals of these functions are easily determined to be
$\frac{\pi}{4}\approx0.8$ and one.  So we have that
$\frac{\pi}{4}\approx0.8\leq F^f_{sym}(\mu)\leq 1$ which is
indeed correct as one can see from the numerical evaluation of
$F^f_{sym}(\mu)$ plotted in figure \ref{integrals.ps}.

As mentioned before we need to introduce a UV cutoff for the bare
charges, allowing the fluxes to approach a charge only up to a minimal
distance $R_0$. One way to put this is that for the symmetrical
process to be able to take place, or for that matter any decay mode
using fluxes, $w$ needs to exceed a minimal value depending on $R_0$
and $\mu$. This constraint on $w$ is easily determined with the help
of formula \ref{SymVNeedle} by putting $b=w-R_0$ in other words $y=
1-r_0$ with $r_0=R_0/w$. Determining the zero of the potential than
gives the minimal value of $w$, yielding:
\begin{equation}
\frac{1}{r_0} = \frac{w}{R_0} \geq 2
e^{\mu/2}\left(e^{\mu/2}+\sqrt{e^{\mu}-1}\right)\quad.
\end{equation}

\noindent\textbf{The asymmetric channel:}

The asymmetric channel is the channel where one of the fluxes stays
close to one of the charges and the other flux moves away. An
interesting fact about this decay channel is, that in the limit of
widely separated charges, $w\to\infty$, this channel will still give a
finite decay time, whereas the symmetric channel would not. The energy
gain due to the presence of a flux pair in this system again follows
from formula \ref{Eneedle}. We fix one of the fluxes at the minimal
cut-off distance $R_0$ from one of the charges.  The other flux is
pushed away from this charge. In this case it is natural to scale the
variables by the core size $R_0$ as this is the only length scale in
the limit of $w\to\infty$, so we define $\tilde{w}=\frac{w}{R_0}$ and
$\tilde{y}=\frac{b}{R_0}$.  The energy gain of this configuration is
given by:
\begin{equation}
E_{fpair}=\frac{2m_f}{\mu}\log\left(\frac{1}{2}\left(1+\frac{2\tilde{w}(1+\tilde{y})-1-2\tilde{y}}{\sqrt{(2\tilde{w}-1)(2\tilde{w}-2\tilde{y}-1)(2\tilde{y}+1)}}\right)\right)\quad.
\end{equation}
The potential is obtained by adding the mass term for the creation of
the two alice fluxes out of the vacuum. The action of the bounce is
determined in the same manner as we did in formula \ref{SSymNeedle},
not only do we have a different potential, we also need to change a
factor $4$ into $2$, because only one flux is moving in this decay
channel. For the action we obtain the following expression:
\begin{equation}
S^f_{asym} = 4~m_f~b_c~F^f_{asym}(\mu,\tilde{w})\quad,
\end{equation}
where the critical separation $b_c$ is given by:
\begin{equation}
b_c=\frac{(2\tilde{w}-1)R_0}{2-2\tilde{w}+\frac{\tilde{w}\left(\cosh\left(\frac{\mu}{2}\right)+3\sinh\left(\frac{\mu}{2}\right)\right)}{\sqrt{e^{\mu}-1}}}\quad.
\end{equation}
The function $F^f_{asym}(\mu,\tilde{w})$ is defined by:
\begin{equation}
F^f_{asym}(\mu,\tilde{w}) = \int^1_0
\sqrt{1-\frac{1}{\mu}\log\left(\frac{1}{2}\left(1+\frac{2\tilde{w}(1+\tilde{y}^\prime \tilde{y}_c)-1-2\tilde{y}^\prime \tilde{y}_c}{\sqrt{(2\tilde{w}-1)(2\tilde{w}-2\tilde{y}^\prime \tilde{y}_c-1)(2\tilde{y}^\prime \tilde{y}_c+1)}}\right)\right)} d\tilde{y}^\prime \quad,
\end{equation}
and depends also on the separation of the external charges
$2\tilde{w}$.  Although we do get a similar expression as in the
symmetric case the integral in the asymmetric case is not that easily
estimated, see figure \ref{integrals2.ps} for a plot of
$F^f_{asym}(\mu,\tilde{w})$ and figure \ref{integrals.ps} for
the value this integral takes in the limit of $\tilde{w}\to\infty$.

The remarkable fact is that this action remains finite in the limit of
$\tilde{w}\to\infty$. Thus the decay time of a single charge (i.e., of
charge itself) is finite in two dimensional alice electrodynamics.

\textbf{The constant field:}

Next we investigate the decay width per volume of a constant electric
field. The energy gain due to the presence of a flux pair in line with
the electric field strength can be found from formula
\ref{SymENeedle}. We move the charges to infinity and increase $Q$
such that the ratio $Q/w$ is kept fixed and we define the electric
field as ${\cal E}\equiv\frac{Q}{\pi w}$. The resulting energy gain
due to the presence of a flux pair then equals:
\begin{equation}
E_{fpair}= \pi {\cal E}^2 b^2\quad.
\end{equation}
The action is easily determined to be:
\begin{equation}
S^f_{const}= \sqrt{2} \pi m_f b_c \quad,
\end{equation}
where the critical separation is,
\begin{equation}
b_c=\sqrt{\frac{2 m_f}{{\cal E}^2 \pi}} \quad.
\end{equation}
The result is of course independent of position as it determines the
decay rate per unit volume of a constant electric field.

\subsubsection{Charge decay time due to creation of point charges}
Let us now investigate the field instability of a pair of external
charges under the creation of two dynamical charges. Since point
charges have a singularity in the field energy at the core we
introduce again a cutoff $R_0$ to regulate some of the infinities in
our calculations. First we will determine the energy gain due to the
presence of two point charges. We denote the two initial charges as
$C_1$ and $C_2$, the created charges as $D_1$ and $D_2$. We put the
four charges on one line and obviously assume the charges to be
alternating. Symbolically the energy gain can be written as:
\begin{eqnarray}
E^q_{gain} &=& (C_1+C_2)^2 - (C_1+C_2+D_1+D_2)^2\\ &=&
-(D_1^2+D_2^2+2D_1D_2) - 2(C_1+C_2)(D_1+D_2)\quad.
\end{eqnarray}
The first part, $D_1^2+D_2^2$, has an infinite contribution at the
cores of the charges and only these infinities will be removed, i.e.,
only in this term we cut away a disc with radius $R_0$ around the
charges. Taking the origin halfway the two created charges and
denoting the distances of the charges with respect to this origin
$w_1$, $w_2$ and $b$, the energy difference $E^q_{gain}$ is
given by:
\begin{equation}
E^q_{gain} = -
\frac{Q^2}{\pi}\log{\left(\frac{2(x_1-1)(x_2-1)}{(x_1+1)(x_2+1)\tilde{r}_0}\right)}\quad,
\end{equation}
with $x=\frac{w}{b}$ and $\tilde{r_0}=\frac{R_0}{b}$.\\ This is the
change in energy due to the electrical field configuration. We still
need to take the mass of the point charges into account. We assume
that the point charges are created a distance $2 R_0$ away from each
other and the energy cost of this process we call $2 m_q$. Thus the
total energy gain is given by:
\begin{equation}
V_{qpair} = 2 m_q\left(1 +
\frac{1}{\nu}\log\left(\frac{(x_1-1)(x_2-1)(x_1+\tilde{r}_0)(x_2+\tilde{r}_0)}{(x_1-\tilde{r}_0)(x_2-\tilde{r}_0)(x_1+1)(x_2+1)\tilde{r}_0}\right)\right)\quad, 
\label{Egainpoint}
\end{equation}
with $\nu=\frac{2\pi m_q}{Q^2}$.\\ Next we will use this energy gain
to determine the action of the bounce in different channels of the
decay process.

\noindent\textbf{The symmetric channel:}

\begin{figure}[!htb]
\begin{center}
\includegraphics[height=12cm,angle=270,clip]{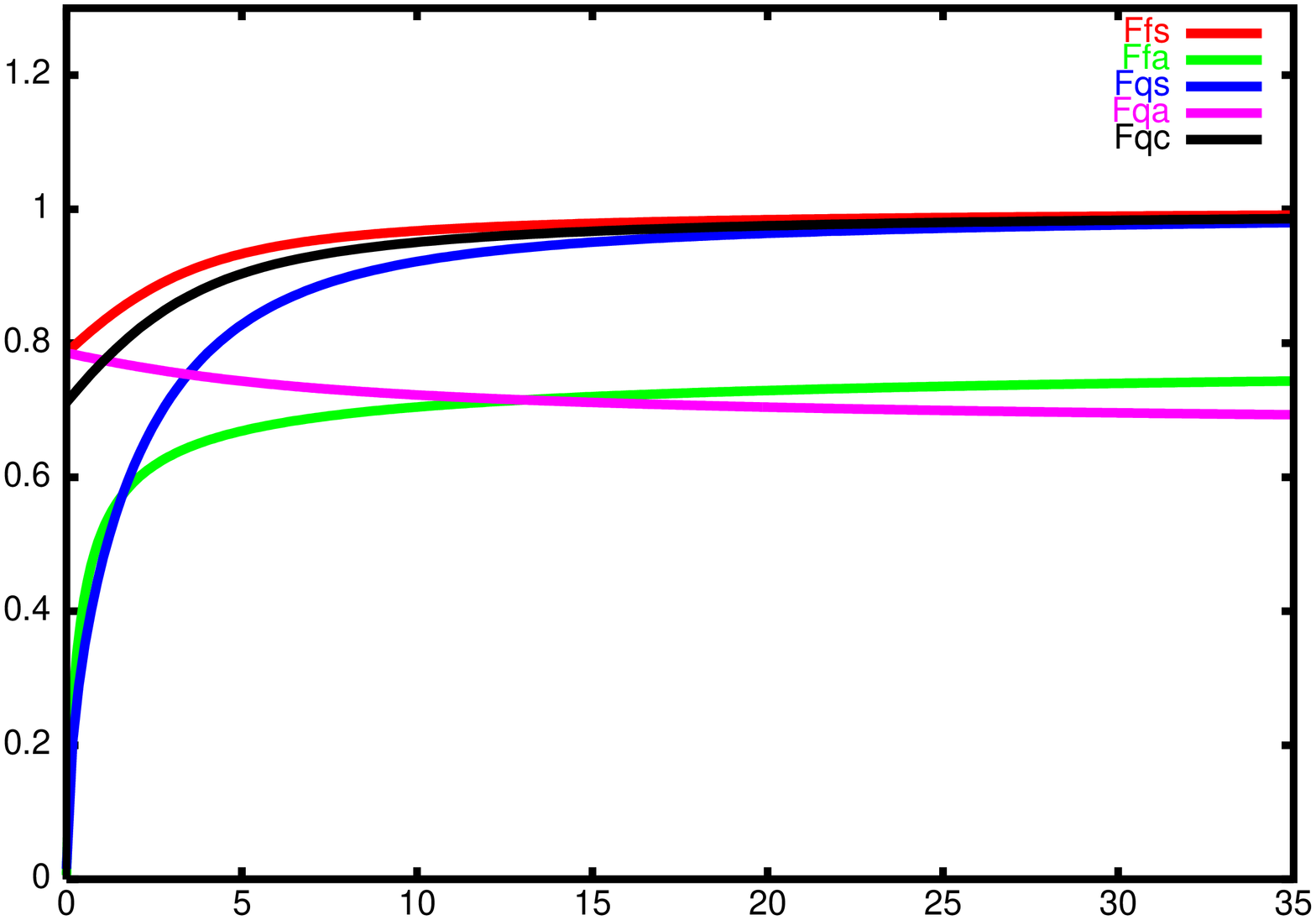}
\caption[somethingelse]{\footnotesize \\This figure shows the five
functions $Ffs=F^f_{sym}(\mu)$, $Ffa=F^f_{asym}(1/\mu)$, $Fqs=(y_c-y_-)F^q_{sym}(\lambda-\lambda_{min})$, $Fqa=F^q_{asym}(\nu)$ and $Fqc=F^q_{const}(\kappa-1)$ numerically, with $\lambda_{min}=\log\left(\frac{1}{2}\left(11+5\sqrt{5}\right)\right)$, $F^q_{asym}(\nu)=F^q_{asym}(\nu,\infty)$ and $F^f_{asym}(1/\mu)=F^f_{asym}(1/\mu,\infty)$}
\label{integrals.ps}
\end{center}
\end{figure}

In the symmetric channel the potential is given by:
\begin{equation}
V_{qpair} = 2 m_q\left(1+
\frac{1}{\nu}\log\left(\frac{(y-1)^2(r_0+1)^2y}{(r_0-1)^2(y+1)^2r_0}\right)\right)\quad,
\end{equation}
where we still use $y=\frac{1}{x}=\frac{b}{w}$ and
$r_0=\frac{R_0}{w}$.\\ To determine the action of the bounce we need
to determine:
\begin{equation}
S_{qpair}=2w\int^{y=y_c}_{y=r_0}\sqrt{4 m_q V_{qpair}}~dy\quad.
\end{equation}
This is a quite non-trivial integral. We will estimate this integral
by slightly changing the boundary conditions. As the lower boundary
condition we will not take $r_0$, but the point between $r_0$ and zero
where $V_{qpair}=0$. Later we will estimate the part we add to the
action by this change in the boundary conditions.\\ So first we will
determine the integral:
\begin{equation}
S^{q,1}_{sym}=2w\int^{y=y_c}_{y=y_-}\sqrt{4 m_q
V_{qpair}}~dy\quad,
\end{equation}
with $y_c$ and $y_-$ the two values of $y$ where $V_{qpair}$ is equal
to zero and with $0<y_-<y_c<1$. This integral is still quite
difficult. We can determine it up to a part that we evaluate
numerically and understand quite well. The action can be written as:
\begin{equation}
S^{q,1}_{sym}= 4\sqrt{2}m_q
(b_c-b_-)\gamma^q_{sym}F^q_{sym}(\lambda)\quad,
\end{equation}
with $\gamma^q_{sym}=\sqrt{\frac{\lambda-\lambda_{min}}{\nu}}$,
$\lambda=\nu+\log\left(\frac{(r_0+1)^2}{(r_0-1)^2 r_0}\right)$,
$\lambda_{min}=\log\left(\frac{1}{2}\left(11+5\sqrt{5}\right)\right)$
and $F^q_{sym}(\lambda)$ given by:
\begin{equation}
F^q_{sym}(\lambda)=\int^{1}_{0}\sqrt{\frac{1+\lambda^{-1}
\log\left(\frac{(1-((y_c-y_-)y^\prime+y_-))^2}{(1+((y_c-y_-)y^\prime+y_-))^2}((y_c-y_-)y^\prime+y_-)\right)}{1-\frac{\lambda_{min}}{\lambda}}}~dy^\prime\quad,
\end{equation}
with $y_c=b_c/w$ and $y_-=b_-/w$.\\ In figure \ref{integrals.ps} we
have plotted a numerical evaluation of the function
$(y_c-y_-)F^q_{sym}(\lambda)$.  We still need to estimate the part
introduced by taking different boundary values. This may be estimated
by the maximum of the integrand in the region between $y_-$ and $r_0$
times $r_0$. If $\left(-2+\sqrt{5}\right)>r_0$ $V_{qpair,max}=2m_q$
else $V_{qpair,max}=2
m_q\sqrt{\frac{\lambda-\lambda_{min}}{\nu}}$. Thus we estimate this
part of the action to be $S^{q,2}_{sym}$, which is typically much
smaller than $S^{q,1}_{sym}$, as follows:
\begin{eqnarray}
S^{q,2}_{sym}&\leq& 4\sqrt{2} m_q R_0\quad
\mbox{if}\quad\left(-2+\sqrt{5}\right)> r_0\\  \mbox{and else}& &\nonumber\\
S^{q,2}_{sym}&\leq& 4\sqrt{2} m_q R_0
\sqrt{\frac{\lambda-\lambda_{min}}{\nu}}\quad.
\end{eqnarray}

\textbf{The asymmetric channel:}

\begin{figure}[!htb]
\begin{center}
\mbox{\psfig{figure=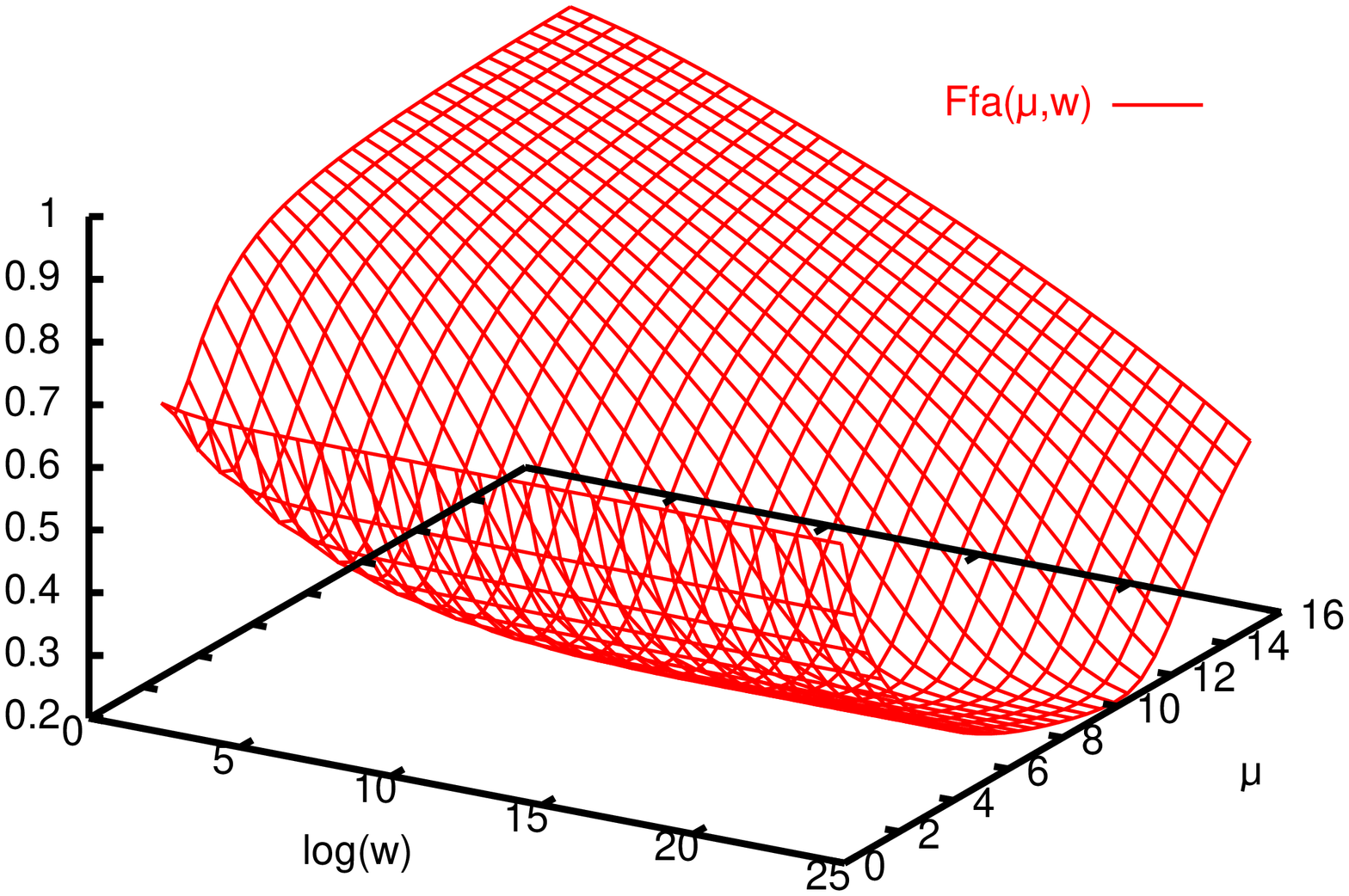,width=7.9cm,angle=270}}
\mbox{\psfig{figure=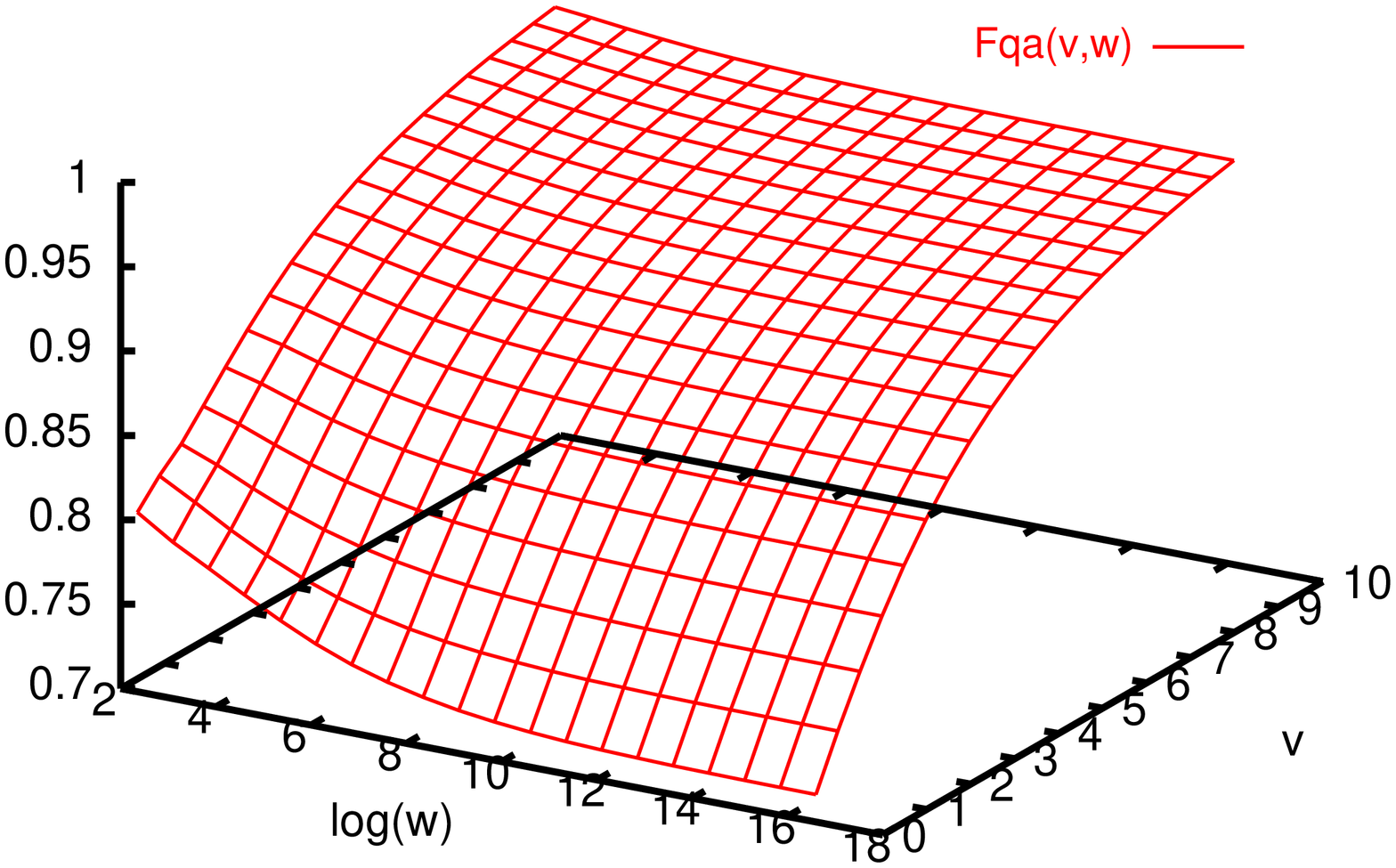,width=7.9cm,angle=270}}
\makebox[7.9cm][c]{\footnotesize{(a)}}
\makebox[7.9cm][c]{\footnotesize{(b)}}
\caption[somethingelse]{\footnotesize\\ Figure (a) shows a plot of
$F^f_{asym}(\mu,\tilde{w})$. The figure shows that the limit of the
integral at $\tilde{w}\to\infty$ is reached only very slowly. The
minimum value of the integral only moves slowly to large $\mu$ as
$\tilde{w}$ grows exponentially. Figure (b) shows a plot of
$F^q_{asym}(\nu,\tilde{w})$. The figure shows that the limit of of the
integral at $\tilde{w}\to\infty$ is reached very fast. In the limit of
$\tilde{w}\to\infty$ we know the integral exactly.}
\label{integrals2.ps}
\label{integrals3.ps}
\end{center}
\end{figure}

In the asymmetric channel the potential is given by:
\begin{equation}
V_{qpair} = 2 m_q\left(1 +
\frac{1}{\nu}\log\left(\frac{2\tilde{y}(\tilde{w}-\tilde{y}-1)}{(\tilde{y}+1)(\tilde{w}-2)}\right)\right)\quad,
\end{equation}
The action of the bounce is given by:
\begin{equation}
S^q_{asym}=4 m_q b_c \gamma^q_{asym}F^q_{asym}(\nu,\tilde{w})\quad,
\end{equation}
with
$\gamma^q_{asym}=\sqrt{1+\frac{1}{\nu}\log\left(\frac{2\left(1-\sqrt{\tilde{w}}\right)^2}{\tilde{w}-2}\right)}$
and
\begin{eqnarray}
b_c&=&R_0\left(\frac{1}{4}e^{-\nu}(2+2e^\nu(\tilde{w}-3)-\tilde{w}+\nonumber\right.\\&&\left.\sqrt{-8e^\nu(\tilde{w}-2)+(-2-2e^\nu(\tilde{w}-1)+\tilde{w})^2}\right)\quad,
\end{eqnarray}
and $F^q_{asym}(\nu,\tilde{w})$ is given by:
\begin{equation}
F^q_{asym}(\nu,\tilde{w})=\int_0^{1}\sqrt{\frac{1+\frac{1}{\nu}\log\left(\frac{2(\tilde{y}^\prime
\tilde{y}_c +1)(\tilde{w}-\tilde{y}^\prime
\tilde{y}_c-2)}{(\tilde{y}^\prime
\tilde{y}_c+2)(\tilde{w}-2)}\right)}{1+\frac{1}{\nu}\log\left(\frac{2\left(1-\sqrt{\tilde{w}}\right)^2}{\tilde{w}-2}\right)}}~d\tilde{y}^\prime.
\end{equation}
In the limit $\tilde{w}\to\infty$ we can determine the integral
exactly, yielding:
\begin{equation}
F^q_{asym}(\nu,\infty)= \frac{4
e^\nu-\frac{\sqrt{\pi}~\mbox{Erfi}[\sqrt{\nu+\log(2)}]}{\sqrt{\nu+\log(2)}}}{4
e^\nu-2}\quad.
\end{equation}
Plots of $F^q_{asym}(\nu,\tilde{w})$ and $F^q_{asym}(\nu,\infty)$ are
given in figures \ref{integrals3.ps} and \ref{integrals.ps}.

We recall that in the asymmetric channel for the creation of alice
fluxes the result remains finite in the limit of widely separated
external charges, obviously this is not the case for the action of the
bounce corresponding to the creation of a pair of point charges.

\textbf{The constant field:}

Finally we will consider the case of a constant electric field and
examine the action of the bounce if two point charges are created. To
determine the energy gain in the field configuration we can use
formula (\ref{Egainpoint}). However we cannot take the charge of the
initial charges equal to the charge of the created point charges. To
get the configuration in a finite electric field we take the distance
between the initial charges to infinity while keeping the charge over
the distance ratio fixed. Again we take the electric field ${\cal
E}=\frac{Q_{initial}}{w\pi}$. The potential for the creation of two
point charges in a constant electric field is given by:
\begin{equation}
V_{qpair} = 2 m_q\left(1 + \frac{1}{\nu}\left(\log(\tilde{y}) -
u(\tilde{y}-1)\right)\right)\quad,
\end{equation}
with $u=\frac{4\pi {\cal E}R_0}{Q^2}$.\\ The action of the bounce is
given by:
\begin{equation}
S^q_{const}=2
R_0\int^{\tilde{y}=\tilde{y}_c}_{\tilde{y}=1}\sqrt{4 m_q
V_{qpair}}~d\tilde{y}\quad.
\end{equation}
Just as in the symmetric channel we take slightly different boundary
conditions and estimate the difference later on. We will use the two
values of $\tilde{y}$ where $V_{qpair}=0$ and we get:
\begin{equation}
S^{q,1}_{const}=2R_0\int^{\tilde{y}=\tilde{y}_c}_{\tilde{y}=\tilde{y}_-}\sqrt{4
m_q V_{qpair}}~d\tilde{y}\quad.
\end{equation}
This leads us to:
\begin{eqnarray}
S^{q,1}_{const}=4 m_q(b_c-b_-)\gamma^q_{const}F^q_{const}(\kappa)\quad,
\end{eqnarray}
with $\gamma^q_{const}=\sqrt{\frac{\kappa-1}{\nu}}$,
$\kappa=\nu-\log(u)+u$,
$b_c\frac{u}{R_0}=\left(\tilde{y}_c^\prime=\right)\kappa+\log(\kappa+\log(\kappa+\log(\kappa+\cdots)))$
and
$b_-\frac{u}{R_0}=\left(\tilde{y}_-^\prime=\right)\exp(-\kappa+\exp(-\kappa+\exp(-\kappa+\cdots)))$,
where $\tilde{y}_c^\prime$ and $\tilde{y}_-^\prime$ are the two real
solutions of
$\kappa+\log(\tilde{y}^\prime)-\tilde{y}^\prime=0$. $F^q_{const}(\kappa)$
is a function which varies only from $\frac{\pi}{4}$ at $\kappa\to 1$
to $\frac{2}{3}$ at $\kappa\to\infty$ and is given by:
\begin{equation}
F^q_{const}(\kappa)=\int^{1}_{0}\sqrt{\frac{\kappa+\log((\tilde{y}_c^\prime-\tilde{y}_-^\prime)\tilde{y}^\prime+\tilde{y}_-^\prime)-((\tilde{y}_c^\prime-\tilde{y}_-^\prime)\tilde{y}^\prime+\tilde{y}_-^\prime)}{\kappa-1}}~d\tilde{y}^\prime
\end{equation}
see figure \ref{integrals.ps} for a plot of $F^q_{const}(\kappa)$.\\
We still need to estimate the part we introduced by taking different
boundary values. We approximate that part by the maximum of the
integrand in the region between $\tilde{y}_-$ and $1$ times $R_0$. If
$u<1$ then $V_{qpair,max}=2 m_q$ and otherwise $V_{qpair,max}=2
m_q\left(\frac{\kappa-1}{\nu}\right)$. Thus the upperbound for this
part of the action, $S^{q,2}_{const}$, is typically much smaller than
$S^{q,1}_{const}$, to be explicit:
\begin{eqnarray}
S^{q,2}_{const}&\leq& 4\sqrt{2} m_q R_0 \quad \mbox{if}\quad u<1\\
& \mbox{else}&\nonumber\\ S^{q,2}_{const}&\leq& 4\sqrt{2} m_q
R_0\sqrt{\frac{\kappa-1}{\nu}}\quad.
\end{eqnarray}

\subsubsection{Comparing the decay channels}
We just determined the actions of bounce solutions corresponding to
some decay channels of two static point charges. As expected the
action depends strongly on the parameters of the model. LAED allows
for the different parameters to be independent of each other, so there
are many possibilities for the preferred decay channel. Although the
LAED model we described before does not require dynamical charges we
did determine the action of some decay channels for the creation of
pairs of such charges. Both dynamical charges and alice fluxes can
render the static point charge configuration unstable. However, the
decay time will typically depend exponentially on the distance between
the two static charges except for one possible mode: the asymmetric
decay channel of the two static charges under the creation of two
alice fluxes. The action of this channel saturates. This means that
the even the decay width of a single point charge is finite in AED,
this obviously in contrast with ordinary ED. This instability is the
process mentioned at the end of section \ref{dipole}, which may be
considered as the two dimensional dual analog of the monopole core
instability described in \cite{jelper3}.  This implies basically the
nonexistence of static charges in the theory, and that is the main
observation we make in this paper.

We already mentioned that a pair of alice fluxes can be represented by
a conducting needle in our configurations. On a conductor charges are
free to move and one can for example have an induced dipole moment. In
this picture the creation of two point charges is just a highly
singular charge distribution on this line segment and it is obvious
that the action of the bounce for alice fluxes can always be made
lower because the charge distribution can still be varied. A simple
and extreme example is the asymmetric channel in the limit of
$\tilde{w}\to\infty$. Here the action of the bounce for the point
charges is infinite while the action of the bounce for the alice
fluxes remains finite.

\section{Conclusions and outlook}
In this paper we have extensively analyzed the behavior of alice
fluxes in the presence of electric charges in (2+1)-dimensions. We
showed that a pair of alice fluxes in the presence of an electric
charge develops an induced electric dipole moment. This dipole moment
is of the cheshire type which means that it is carried by the flux
pair, and that the would-be charges making up the dipole are strictly
nonlocalizeable en thus remain elusive. Exploiting conformal invariance
we determined the resulting field configurations exactly which in turn
allowed us to calculate the energy gain due to the introduction of a
pair of alice fluxes between two external charges. Subsequently we
considered the stability using semi-classical methods, using a
Euclidean bounce solution.  

\begin{figure}[!htb]
\begin{center}
\includegraphics[height=11.5cm,angle=270,clip]{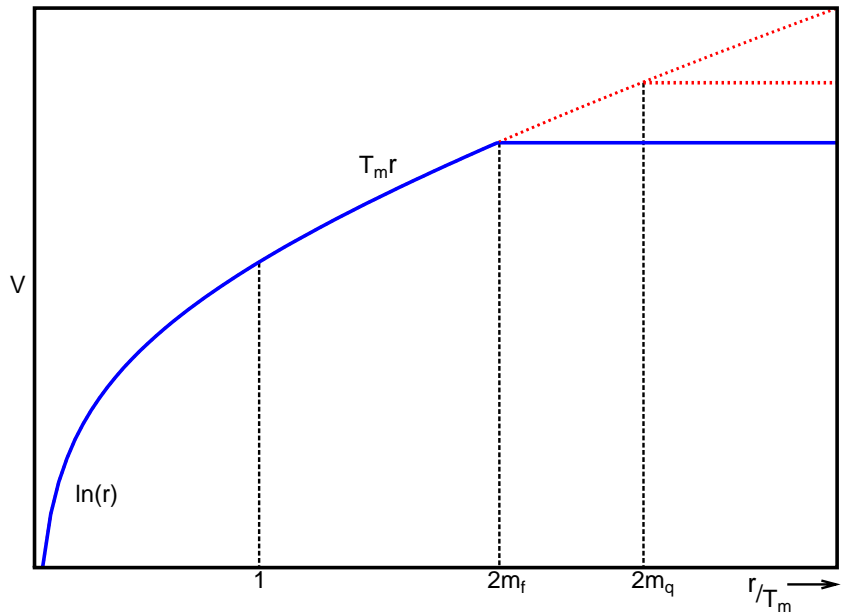}
\caption[somethingelse]{\footnotesize  The effective potential for a
pair of external charges. The figure represents the case where the
mass of the alice flux pair is larger than the confinement scale, but
smaller than the mass of dynamical charges. It is possible to lower
$m_f$ below the confinement scale.}
\label{potential.eps}
\end{center}
\end{figure}

We used a lattice model of AED \cite{jelper4} to investigate the
effects of alice fluxes on a configuration of static point charges,
because it allowed us to investigate the effects of the different
topological defects separately. In the case of heavy monopoles we
found an instability in the charge configuration due to the creation
of a pair of alice fluxes. Although this instability looks quite
similar to the instability due to the creation of two dynamical point
charges there is a crucial difference. In the limit of increasing
separation between the static charges the decay time due to the
creation of dynamical point charges diverges, while for the creation
of two alice fluxes it saturates and remains finite. To reach this
conclusion we did not have to calculate the fluctuation determinant in
detail, assuming that it is finite. Consequently in (L)AED a single
bare charge is unstable under the creation of two alice fluxes, which
can be seen as the (2+1)-dimensional dual analog of the monopole core
instability~\cite{jelper3}. If the monopole mass moves down, i.e., the
confinement scale comes into play, the instabilities due to a flux
pair and a charge anti-charge pair become very similar. In figure
\ref{potential.eps} we have sketched the potential for a typical
situation.

For the theory at hand we are presently determining
these potentials through computer simulations in the lattice alice
electrodynamics model \cite{jelper4} mentioned at the beginning of
this paper and hope to report on these results in the near future.

Let us now give some comments on the continuum theory. We expect
the situation to be not so much different. The topological defects
arise as a consequence of spontaneous symmetry breaking, which means
that the mass scales for the fluxes and monopoles might be much more
constrained. We showed~\cite{jelper3} that if the flux mass gets much
less then the monopole mass, one may well get that the monopole decays
in a flux ring carrying a cheshire magnetic charge. This suggests that
the confinement scale and the instability scale (due to flux creation)
cannot be too much different.  As we explained, if the monopole and
alice flux mass are comparable the potential still saturates due to
the instability under the creation of two alice fluxes. 

In this paper we showed that the possibility of cheshire charge in a
theory has serious consequences for the stability of charge in the
theory in two dimensions. It is usually a question of energetics what
the stable configuration is, but for theories which allow for cheshire
charges, a cheshire charge configuration is the natural second
candidate to carry the charge. This suggests that any theory which
breaks to a subgroup which contains a discrete and continuous
component that do not mutually commute the gauge charges may well
become unstable due to the cheshire phenomenon.  Another interesting
class of theories which typically contain cheshire charged
configurations are the theories with non-abelian discrete gauge
symmetries, which are best described with the help of a spontaneously
broken Hopf symmetry \cite{slinger1,slinger2}.

In the appendix of this paper we introduce an object called the
(magnetic) cheshire current and we discuss its relation with
(electric) cheshire charges. We will also discuss its relation with
the closed electric field lines that occur if one interprets the
occurrence of an instanton as an event in the (2+1)-dimensional
(alice) electrodynamic setting. From this picture the confinement
mechanism can be understood quite easily.

Acknowledgment: We thank Jan Smit for very useful discussions on the
topics discussed in this paper. This work was partially supported by
the ESF COSLAB program.

\section*{Appendix}
\appendix
\section{Cheshire current and confinement}

In this appendix we will discuss the notion of a (magnetic) cheshire
current in AED and the confinement of charges in (2+1)-dimensional
(alice) electrodynamics \cite{Polyakov2}. We'll introduce a
configuration in AED named (magnetic) cheshire current and explain its
relation with (electrical) cheshire charges and confinement in two
dimensions. We'll introduce a picture of two dimensional confinement
from which qualitatively the confinement of the electrical flux into a
flux tube comes apparent.

\subsection{The cheshire current}
\label{Chescurrent}\label{nlcurrent}
Neither electric nor magnetic field lines are allowed to cross an alice
flux, suggesting some exotic type of super conductivity through the
core of the flux tube. In this part of the appendix we return to this
analogy and find an interesting gauge complementarity between electric
cheshire charges and a magnetic cheshire currents. Let us introduce
the latter first.\\[2mm] Let us consider the following ``gedanken''
experiment. We create two charged particles from the vacuum and take
one of the two particles around two spatially separated fluxes and
then annihilate the two particles again. If the flux tubes are
magnetic super-conductors this would have resulted in two magnetic
current carrying fluxes, each with closed electric field lines around
them. In the case of two alice fluxes a different picture
emerges. Since the field lines cannot close around a single alice flux,
one needs to take an even number of fluxes to be able to annihilate
the particles again. This means that if one pulls the two fluxes apart
one cannot be left with two fluxes which each carry a current. The
field lines need to stay around both fluxes. A situation very different
from the super conductors indeed.  The system as a whole carries the
current and just as in the case of a cheshire charge the current is
non-localizeable; we should call this object a cheshire current.

\begin{figure}[!htb]
\begin{center}\mbox{\psfig{figure=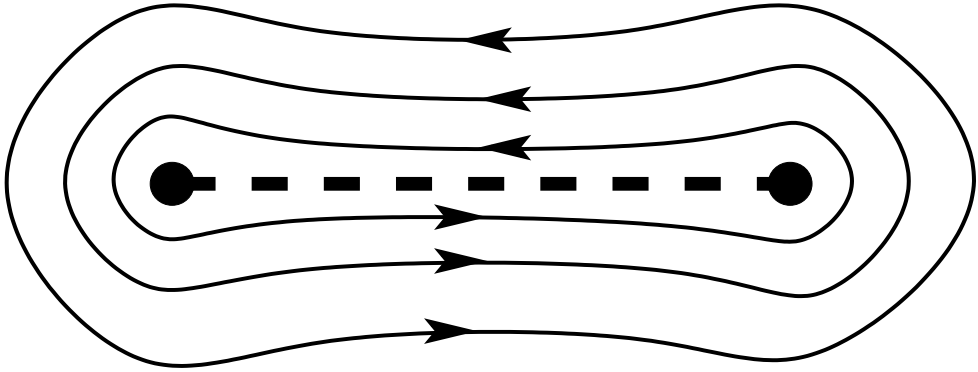,width=13cm}}
\caption[somethingelse]{\footnotesize Closed electric field lines of a
(magnetic) cheshire current configuration.}
\label{current3.eps}
\end{center}
\end{figure}

The resulting field line configuration, depicted in figure
\ref{current3.eps}, implies an attractive interaction between the two
fluxes, on top of the normal flux interactions. It has the opposite
effect of a cheshire charge, which leads to a repulsive force between
the two fluxes.

\begin{figure}[!htb]
\begin{center}
\mbox{\psfig{figure=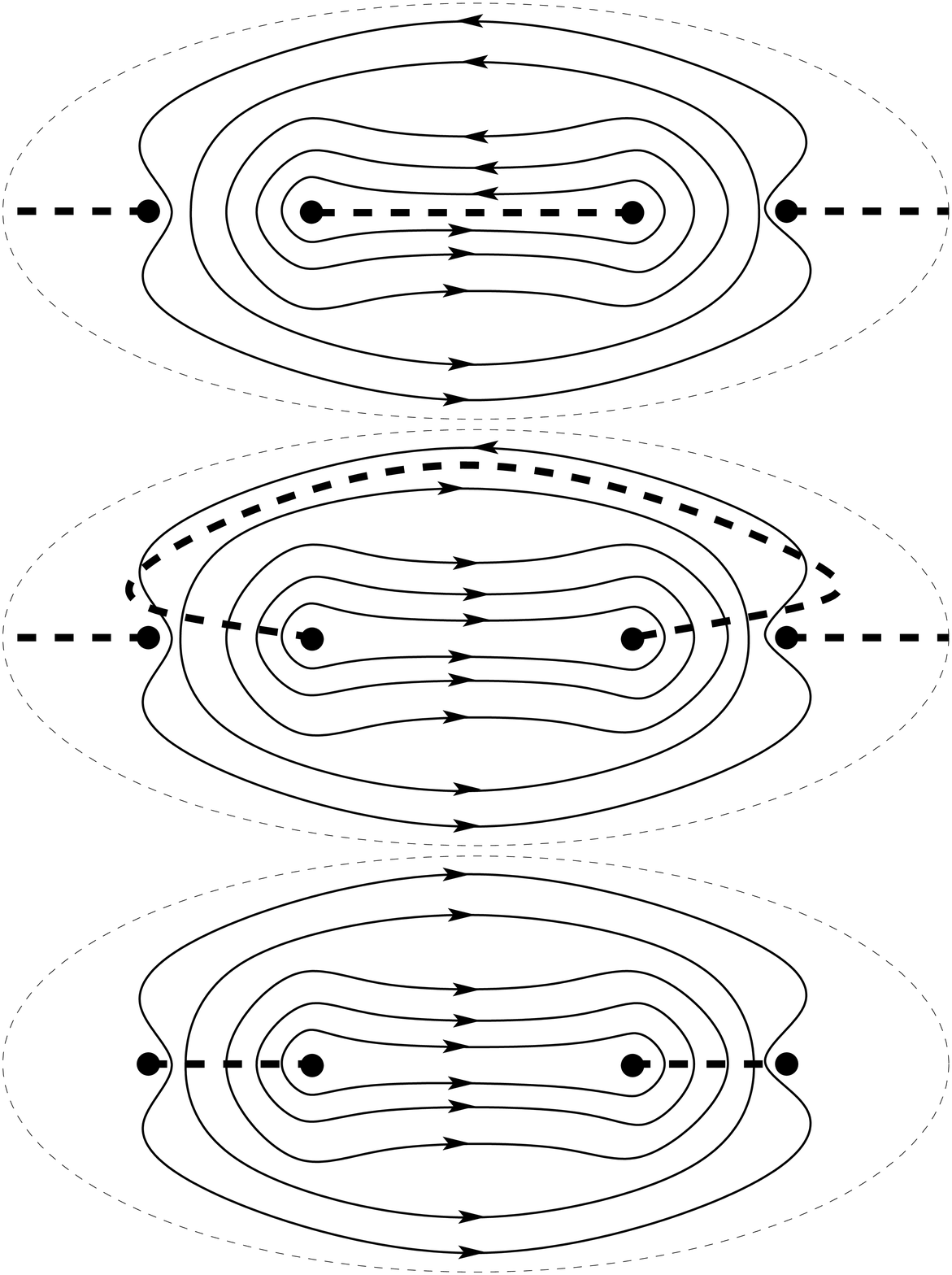,width=11cm}}
\caption[somethingelse]{\footnotesize The 'duality' transformation
from three magnetic cheshire currents into two electric cheshire
charges. }
\label{confinement.eps}
\end{center}
\end{figure}

Upon closer inspection we will see that there is a certain {\em gauge
complementarity}, reconciling the two different pictures, describing
non-localizeable alice effects. At first sight electric cheshire
charge and a magnetic cheshire current appear to be very different
entities. Let us now point out that there is actually a close relation
between them. Imagine we repeat the gedanken experiment we just
performed, but now we move in two more alice fluxes from infinity in
such a way that all four of them are on one single line. As we know,
on each flux one $\ZZ_2$ line should end. For convenience we put these
half lines on top of the line on which we put the fluxes. For every
flux we then still have the freedom to let the line go to the left or
to the right.  The result just yields two different, but gauge
equivalent, configurations, as is illustrated by the top and bottom
pictures in figure \ref{confinement.eps}.

\noindent As we argued before, one can deform the $\ZZ_2$ lines in any
way one wants by gauge transformations. From figure
\ref{confinement.eps} it is clear that we can gauge transform the
first configuration into the last one. This means that they both
describe the same physics, although their interpretation appears to be
quite different. In one case, see the bottom picture of figure
\ref{confinement.eps}, one would argue that two cheshire charges are
the source of the field lines, but in the other situation, see the top
picture of figure \ref{confinement.eps}, one would argue that three
cheshire currents are the source of the field lines. Apparently there
are two different ways of looking at this configuration. As was
explained before \cite{schwarz} one needs to cut away some region(s)
of space-time if one wants to consider field strengths which are not
single valued in the presence of an alice flux. However, there is of
course not unique choice to do this. This freedom of choice
corresponds exactly to the gauge complementarity of cheshire charge
and cheshire current.

We do note that although they are related by a gauge
transformations it does not mean that all configurations can be
thought of as consisting only of cheshire charges or only of cheshire
currents. A simple example is a pair of alice fluxes carrying a
cheshire charge and a cheshire current. This object may in fact be a
stable configuration in two dimensions, since the electric cheshire
charge results in a repulsive force between the two fluxes whereas the
magnetic cheshire current results in a attractive force between the
two fluxes. These could be made to cancel leading to a stationary
configuration.

\subsection{Confinement in a two dimensional picture}

\begin{figure}[!htb]
\begin{center}
\makebox[7.9cm]{\psfig{figure=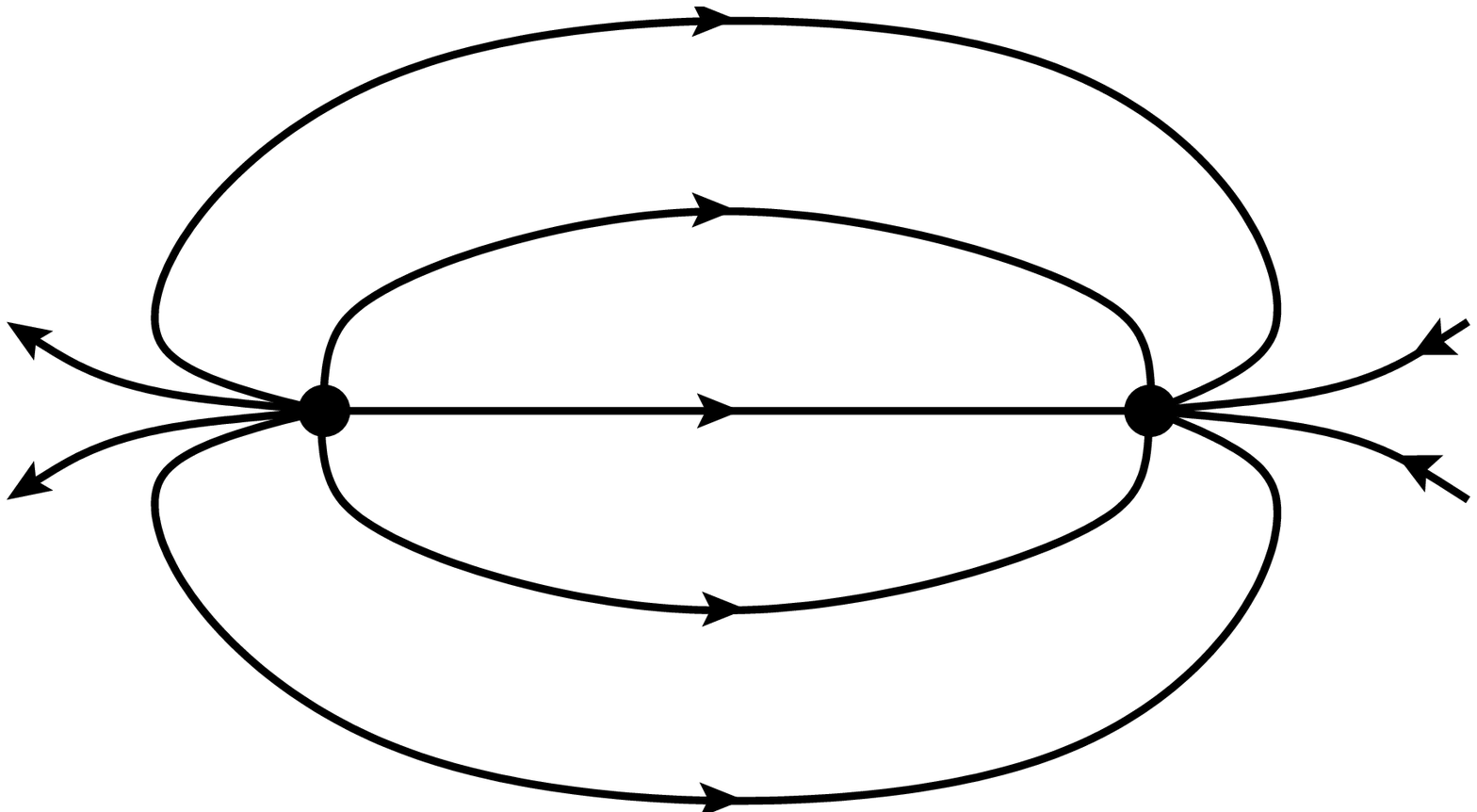,width=7cm}}
\makebox[7.9cm]{\psfig{figure=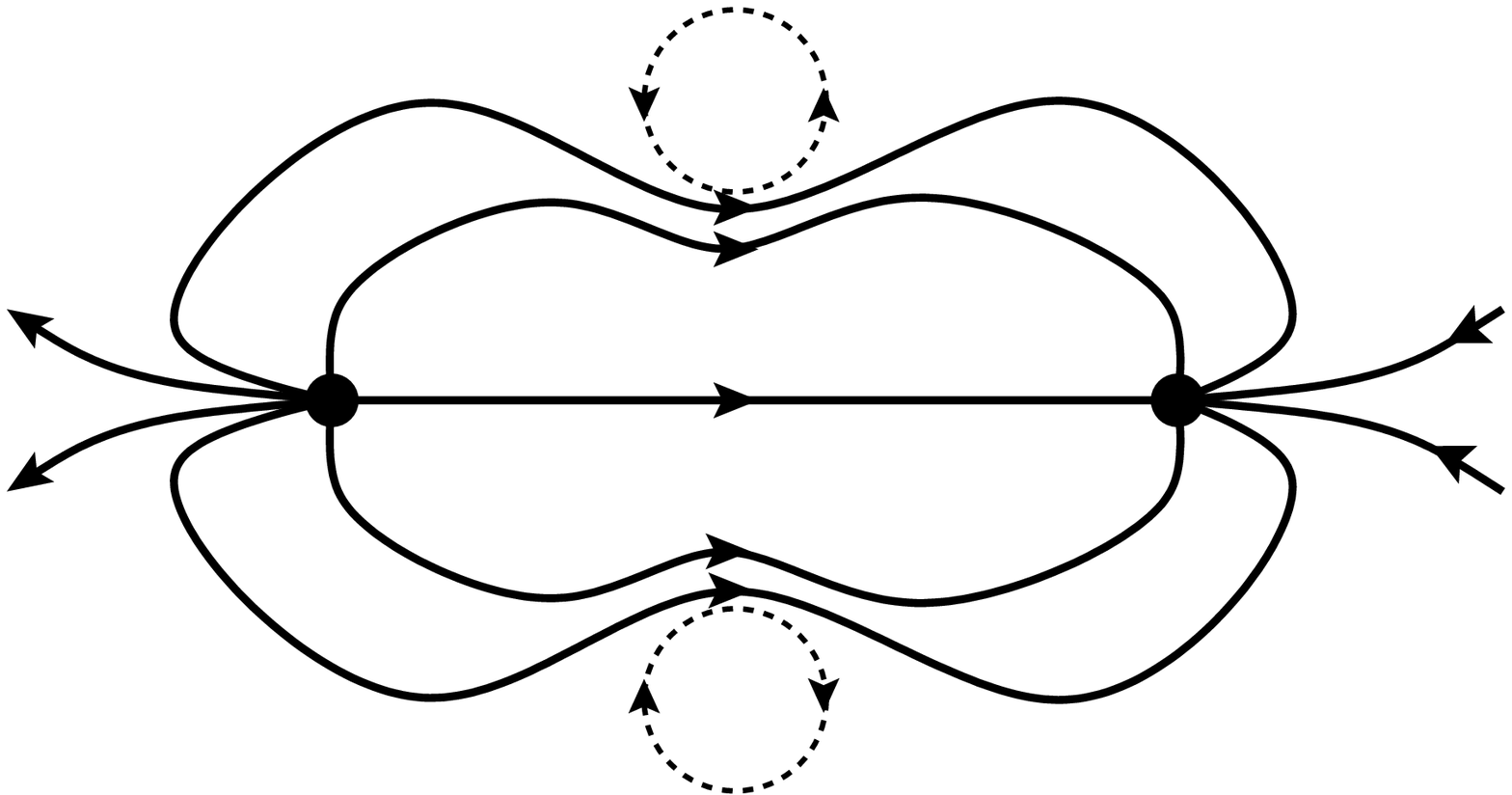,width=7cm}}
\makebox[7.9cm][c]{\footnotesize{(a)}}
\makebox[7.9cm][c]{\footnotesize{(b)}}
\caption[somethingelse]{\footnotesize In figure (a) we plotted the
field configuration of two opposite charges in the absence of
instantons. In figure (b) we see that the introduction of magnetic
currents, representing the instantons in Minkovski space, pushes the
field lines inwards explaining the fact that the electric flux gets
confined in a flux tube in the presence of a (polarized) instanton
density.}
\label{confine.eps}
\end{center}
\end{figure}

In this subsection we will consider the confinement of
(2+1)-dimensional electrodynamics. This problem was already solved in
\cite{Polyakov2}. For any non-zero value of the gauge coupling
constant (2+1)-dimensional electrodynamics is confining (in the
quenched approximation). It is well known that the instanton density
increases and polarizes around the minimal sheet bounded by a closed
Wilson loop. In a three dimensional Euclidean space the instanton
configuration is in fact just a magnetic monopole. After translating
the instanton configuration to Minkovski space it is easy to
understand that the polarization of the instanton density results in
the confinement of the electrical flux into a flux tube.\\ By going to
Minkovski space the interpretation of the fields change. The
$z$-component of the magnetic field becomes the pseudo scalar magnetic
field in the (2+1)-dimensional Minkovski space, while the $\theta$ and
$\rho$ components of the magnetic field get translated into the $\rho$
and $\theta$ components of the electric field respectively. For the
moment we will ignore the factors of $i$ as they will have no
influence on the picture we use, although they do play an important
role in the dynamics and the polarization of the instanton density.\\
Changing from Euclidean to Minkovski space allows us to interpreted
the instanton density as a magnetic current density in Minkovski
space. The nice thing of this two dimensional interpretation is that
the confinement of the electrical flux into a flux tube easily follows
from the superposition of the field lines of the pair of charges and
the magnetic currents. In figure \ref{confine.eps} we see that
superimposing a magnetic current to the electric dipole configuration
moves the field lines inwards. Indicating that a (polarized) magnetic
current density would confine the electric flux into a flux tube.

In the previous section of this appendix we introduced an object in
AED which can also be identified as a magnetic (cheshire)
current. However the dynamics, due to the factors of $i$, is very
different.


\end{document}